\documentclass[aps,superscriptaddress,tightenlines,nofootinbib]{revtex4}
\usepackage{graphicx,amssymb,amsbsy,bm,amsmath}
\usepackage{hyperref}

\newcommand{\vx}{\ensuremath{\vec{x}}}
\newcommand{\vk}{\ensuremath{\vec{k}}}
\newcommand{\vq}{\ensuremath{\vec{q}}}

\newcommand{\bnu}{\ensuremath{\bar{\nu}}}

\newcommand{\be}{\begin{equation}}
\newcommand{\ee}{\end{equation}}
\newcommand{\bea}{\begin{eqnarray}}
\newcommand{\eea}{\end{eqnarray}}
\begin{document}
\title{Particle decay during inflation: \\
 self-decay of inflaton quantum fluctuations  during slow roll. }
\author{D. Boyanovsky}
\email{boyan@pitt.edu} \affiliation{Department of Physics and
Astronomy, University of Pittsburgh, Pittsburgh, Pennsylvania
15260, USA} \affiliation{Observatoire de Paris, LERMA. Laboratoire
Associ\'e au CNRS UMR 8112.
 \\61, Avenue de l'Observatoire, 75014 Paris, France.}
\affiliation{LPTHE, Universit\'e Pierre et Marie Curie (Paris VI)
et Denis Diderot (Paris VII), Laboratoire Associ\'e au CNRS UMR 7589,
Tour 24, 5\`eme. \'etage, 4, Place
Jussieu, 75252 Paris, Cedex 05, France}
\author{H. J. de Vega}
\email{devega@lpthe.jussieu.fr} \affiliation{LPTHE, Universit\'e
Pierre et Marie Curie (Paris VI) et Denis Diderot (Paris VII),
Laboratoire Associ\'e au CNRS UMR 7589,
Tour 24, 5\`eme. \'etage, 4, Place Jussieu, 75252 Paris, Cedex 05,
France}\affiliation{Observatoire de Paris, LERMA. Laboratoire
Associ\'e au CNRS UMR 8112.
 \\61, Avenue de l'Observatoire, 75014 Paris, France.}
\affiliation{Department of Physics and Astronomy, University of
Pittsburgh, Pittsburgh, Pennsylvania 15260, USA}
\author{N. G. Sanchez}
\email{Norma.Sanchez@obspm.fr} \affiliation{Observatoire de Paris,
LERMA. Laboratoire Associ\'e au CNRS UMR 8112.
 \\61, Avenue de l'Observatoire, 75014 Paris, France.}
\date{\today}
\begin{abstract}
Particle decay during inflation is studied by implementing a
dynamical renormalization group resummation combined with a small
$\Delta$ expansion. $\Delta$ measures the deviation from the scale
invariant power spectrum and regulates the infrared. In slow roll
inflation, $\Delta$ is a simple function of the slow roll
parameters $ \epsilon_V, \; \eta_V$. We find that quantum
fluctuations can \emph{self-decay} as a consequence of the
inflationary expansion through processes which are forbidden in
Minkowski space-time.  We compute the \emph{self-decay} of the
inflaton quantum fluctuations
 during slow roll inflation. For wavelengths deep inside the
Hubble radius the decay is \emph{enhanced} by the emission of
ultrasoft collinear quanta, i.e. \emph{bremsstrahlung radiation of
superhorizon quanta} which becomes the leading decay channel for
physical wavelengths $H\ll k_{ph}(\eta)\ll H/(\eta_V-\epsilon_V)$.
The decay of short wavelength fluctuations hastens as the physical
wave vector approaches the horizon. Superhorizon fluctuations decay
with a power law $\eta^{\Gamma}$ in conformal time where in terms of
 the amplitude of curvature perturbations $
\triangle^2_{\mathcal{R}} $, the scalar spectral index $n_s$, the
 tensor to scalar ratio $r$ and slow roll parameters :
$$\Gamma \simeq \frac{32 \; \xi^2_V \;
\triangle^2_{\mathcal{R}}}{(n_s-1+\frac{r}{4})^2}
\left[1+\mathcal{O}(\epsilon_V,\eta_V)\right] \; .
$$
The behavior of the growing mode $ {\eta^{\eta_V-\epsilon_V
+\Gamma}}/{\eta} $ features an anomalous scaling dimension $\Gamma$.
We discuss the  implications of these results for scalar and tensor
perturbations as well as for non-gaussianities in the power
spectrum. The recent WMAP data suggests $\Gamma \gtrsim 3.6 \times
10^{-9} $.
\end{abstract}


\maketitle

\section{Introduction}
A period of accelerated expansion in the early universe, namely
inflation, is nowadays part of standard cosmology since explains the
homogeneity, isotropy and flatness of the observed Universe
\cite{guth}-\cite{liddle}. At the same time, inflation provides a
mechanism for generating metric fluctuations which seed large scale
structure: during inflation physical scales grow faster than the
Hubble radius but slower than it during both radiation or matter
domination eras, therefore physical wavelengths cross the horizon
(Hubble radius) \emph{twice}. Quantum fluctuations generated during
inflation with wavelengths smaller than the Hubble radius become
classical and are amplified upon first crossing the horizon. As they
re-enter the horizon during the decelerated stage these fluctuations
provide the seed for matter and radiation inhomogeneities which
generate structure upon gravitational collapse
\cite{mukhanov,hawking,guthpi,starobinsky,bardeen,bran}. Most of the
inflationary models predict fairly generic features: a gaussian,
nearly scale invariant spectrum of adiabatic  scalar and tensor
primordial perturbations (gravitational waves). These generic
predictions are in spectacular agreement with Cosmic Microwave
Background (CMB) observations. Gaussian \cite{komatsu} and adiabatic
nearly scale invariant primordial fluctuations \cite{spergel}
provide an excellent fit to the WMAP data as well as to a variety of
large scale structure observations. Perhaps the most striking
confirmation of inflation as the mechanism for generating
\emph{superhorizon} (`acausal') perturbations is the anticorrelation
peak in the  temperature-polarization (TE) angular power spectra at
$l\sim 150$ corresponding to superhorizon scales
\cite{kogut,peiris}. The anticorrelation between the E-mode (parity
even) polarization fluctuation and the temperature fluctuation is a
distinctive feature of superhorizon adiabatic fluctuations
\cite{sperzalda}: the (peculiar) velocity gradient generates a
quadrupole temperature anisotropy field around electrons which in
turn generates an E-polarization mode. By continuity, the gradient
of the peculiar velocity field is related to the time derivative of
the density (temperature) fluctuations, hence  the peculiar velocity
and the initial (adiabatic) contribution to the (acoustic)
oscillations of the photon baryon fluid are out of phase by $\pi/2$
\cite{sperzalda}. Thus, the product of these two terms gives an
anticorrelation peak at $k \, c_s \; \eta_{dec} =3 \, \pi/4$ which
corresponds to superhorizon wavelengths since the size of the
horizon is $\sqrt{3}$ larger than the size of the sound horizon. The
WMAP (TE) data \cite{kogut,peiris} clearly displays the
anticorrelation peak at $l \sim 150$ providing perhaps one of the
most striking confirmations of adiabatic superhorizon fluctuations
as predicted by inflation. While the robust predictions of a generic
inflationary model seem to provide an excellent fit to the WMAP
data, different models predict slight differences. Therefore,
theoretical differences between different models as well as
potential experimental deviations from the most generic features are
the focus of intense study. With the ever increasing precision of
CMB observations it is conceivable that forthcoming observations
will allow a narrower determination of inflationary models. Relevant
discriminants between models are: non-gaussianity, a running
spectral index either for scalar and/or tensor perturbations, an
isocurvature component of primordial fluctuations, etc. Already WMAP
reports a hint of running spectral index of scalar perturbations
from the blue on large scales to the red on small scales
\cite{peiris}. Quantum effects associated with interactions can
potentially lead to non-gaussian
correlations\cite{allen}-\cite{bartolo}. Therefore the detection of
a running index (as hinted in the WMAP data) or small
non-gaussianities in the temperature correlations imply potentially
interesting quantum phenomena during the inflationary stage that was
imprinted on superhorizon scales.

The inflaton is usually studied as a homogeneous  classical scalar
field\cite{kolb,coles,liddle}. However, important aspects of the
dynamics require a full quantum treatment for their consistent
description. The quantum dynamics of the inflaton is
systematically treated within a non-perturbative framework and
some consequences on the CMB anisotropy spectrum were analyzed in
ref.\cite{cosmo}.

\medskip

In this article we study \emph{quantum} phenomena during inflation
which contribute to relevant observables in the CMB anisotropies
and polarization.  In particular, we focus on {\it inflaton decay}
during inflation as a potential source of quantum phenomena
contributing to deviations from  scale invariance in the
primordial power spectrum and/or to non-gaussian features. If the
inflaton couples to other particles, then its quantum fluctuations
which seed scalar density perturbations also couple to these other
fields. Consequently, the \emph{decay} of the amplitude of the
\emph{quantum fluctuations} of the inflaton may lead to a
modification of the power spectrum of density perturbations. The
same coupling that is responsible for the decay of the  inflaton
quantum fluctuations can be also
 the source of non-gaussian correlations.

Particle decay is a distinct feature of interacting quantum field
theories and is necessarily an important part of the inflationary
paradigm: the decay of the inflaton into lighter particles
\emph{after inflation} may yield to the radiation dominated stage.
Recently, inflaton decay during a post-inflationary stage has been
considered as a possible source of metric perturbations arising
from an inhomogeneity in the inflaton coupling \cite{7L}. Inflaton
decay has also  been studied as a dissipative mechanism in the
dynamics of the inflaton \cite{berera}, however these studies only
apply when the expansion rate is much smaller than the typical
mass scales.

\medskip

In a previous article \cite{desiterI} we introduced and
implemented a systematic program to study the relaxational
dynamics and particle decay in the case of a rapidly expanding
inflationary stage. Whereby rapid expansion refers to  the Hubble
parameter during inflation being  much larger than the mass of the
particles. In the case of the inflaton, this is the situation of
relevance for slow-roll inflation and a necessary (although not
sufficient) condition for an almost scale invariant power spectrum
of scalar fluctuations \cite{linde,kolb,liddle,lidsey}. In
ref.\cite{prem} inflaton decay was studied in some particular
cases for which a solution of the equations of motion was
available. The method of ref.\cite{prem} was recently applied to
the study of the decay of the inflation in alternative de Sitter
invariant vacua\cite{rich}.

The Minkowski space-time computation of the decay rate is not
suitable for  a stage of rapid expansion (as quantified above):
the rapid expansion of the Universe and the manifest lack of a
global time-like Killing vector allow processes that would be
forbidden by energy conservation in Minkowski space-time. As
emphasized in \cite{woodard,desiterI}, the lack of energy
conservation in a rapidly expanding cosmology requires a different
approach to study particle decay. The correct decay law follows
from the  relaxation in time of the expectation value of the field
out of equilibrium. The  relaxation of the non-equilibrium
expectation value of the field is computed in ref.\cite{desiterI}
using the dynamical renormalization group (DRG) which allows to
extract the decay \emph{law} directly from the real time equations
of motion. The reliability and predictive power of the DRG has
been tested for a wide range of physical situations including hot
and dense plasmas in and out of equilibrium \cite{DRG}.

\medskip

 {\bf The goals of this article:}
We compute the particle decay
of quantum fields minimally coupled to gravity with masses $M$ much
smaller than the Hubble parameter, which is the
 relevant case  for slow roll inflation. This entails a much stronger
infrared behavior than for massless particles conformally coupled to gravity.
The emergence of infrared divergences in
quantum processes with gravitons during de Sitter inflation has been
the focus of a thorough study \cite{IRcosmo,dolgov}. As we will see
 in detail below, similar strong infrared behavior enters in the decay
 of minimally coupled particles with masses $M$ much smaller than the
 Hubble parameter $H$. When $M<<H$ there is a small parameter
 $\Delta \sim M^2/H^2$ which regulates the infrared behavior in de
 Sitter inflation. We find that a similar parameter  $\Delta$ exists in
 quasi de Sitter  slow roll inflation which is a simple function of the
slow-roll parameters. $\Delta$ regulates the infrared in the
self-energy corrections \emph{even for massless  particles} (gravitons).

 \medskip

We begin by studying the general case of a cubic interaction of
scalar particles minimally coupled to gravity,  allowing the decay
of one field  into two others during de Sitter inflation. The
masses of all particles are much smaller than the Hubble constant,
which leads to a  strong infrared behavior  in the self-energy
loops. We introduce an expansion in terms of a small parameter
$\Delta$ which regulates the  infrared and which in the case of de
Sitter inflation is determined by the ratio of the mass of the
particle in the loop to the the Hubble constant. Long-time
divergences associated with secular terms in the solutions of the
equations of motion are systematically resummed by implementing
the DRG introduced in refs.\cite{desiterI,DRG} and lead to the
decay law. We then apply these general results to the case of
quasi-de Sitter slow roll inflation. We show that in this case a
similar small parameter $\Delta$ emerges which is a simple
function of slow-roll parameters and regulates the infrared
behavior {\it even} for massless particles. We study the decay of
superhorizon fluctuations as well as of fluctuations with
wavelengths deep inside the horizon. A rather striking aspect is
that a particle {\bf can decay} into
 \emph{itself} precisely as a consequence of the lack of energy
 conservation in a rapidly expanding cosmology. We then focus on
 studying the decay of the  inflaton quantum fluctuations into
 their \emph{own quanta}, namely the {\it self-decay} of the inflaton
 fluctuations, discussing the potential implications on the power
 spectra of primordial perturbations and to non-gaussianity.

 \vspace{2mm}

 {\bf Brief summary of results:}
\begin{itemize}
\item{In the case of de Sitter inflation for particles with mass
$M\ll H$ a small parameter $\Delta\sim M^2/H^2$ regulates the
infrared. We introduce an expansion in this small parameter $\Delta$
akin to the $\varepsilon$  expansion in dimensionally regularized
critical theories. We obtain the decay laws in a $\Delta$ expansion
after implementing the DRG resummation.}

\item{Minimally coupled particles decay \emph{faster} than those
conformally coupled to gravity due to the strong infrared behavior
both for superhorizon modes as well as for modes with wavelengths
well inside the Hubble radius.}

\item{The decay of short wavelength modes, those inside the
horizon during inflation, is \emph{enhanced} by soft collinear
\emph{bremsstrahlung radiation of superhorizon quanta} which
becomes the dominant decay channel when the physical wave vector
obeys, \be k_{ph}(\eta) \lesssim \frac{H}{\eta_V-\epsilon_V}\; ,
\ee where $\eta_V,\epsilon_V$ are the standard potential slow roll
parameters. }

\item{An  expanding cosmology allows processes that are forbidden
in Minkowski space-time by energy
conservation\cite{woodard,desiterI}: in particular, for masses
$\ll H$, \emph{kinematic thresholds} are absent allowing a
particle to decay into \emph{itself}. Namely, the
\emph{self-decay} of quantum fluctuations is a feature of an
interacting theory in a rapidly expanding cosmology. A
self-coupling of the inflaton leads to the self-decay of its
quantum fluctuations both for modes inside as well as
\emph{outside} the Hubble radius. }

\item{The results obtained in de Sitter directly apply to the
\emph{self decay} of the  quantum fluctuations of the inflaton
during slow roll (quasi de Sitter) expansion. In this case,
$\Delta$ is a simple function of the slow roll parameters. For
superhorizon modes we find that the amplitude of the  inflaton
quantum fluctuations relaxes as a power law $\eta^{\Gamma}$ in
conformal time. To lowest order in slow roll, we find $ \Gamma $
completely determined by slow roll parameters and the amplitude of
the power spectrum of curvature perturbations $
\triangle^2_{\mathcal{R}} $: \be \Gamma = \frac{8 \; \xi^2_V \;
\triangle^2_{\mathcal{R}}}{(\epsilon_V-\eta_V)^2}
\left[1+\mathcal{O}(\epsilon_V,\eta_V)\right]
 \ee
\noindent where $\eta$ is conformal time and
$\xi_V,\eta_V,\epsilon_V$ are the standard slow roll parameters.
As a consequence, the growing mode which dominates at late time
evolves as \be \frac{\eta^{\eta_V-\epsilon_V +\Gamma}}{\eta} \;.
\ee featuring an {\it anomalous dimension} $\Gamma$ slowing down
the growth of the dominant mode.

 The decay of the inflaton quantum fluctuations with wavelengths
 deep within the Hubble radius during slow roll inflation is
 {\bf enhanced} by the infrared behavior associated with the collinear
 emission of ultrasoft  quanta, namely \emph{ bremsstrahlung
radiation of superhorizon fluctuations}. The decay  hastens as the
physical wavelength approaches the horizon because the phase space
for the emission of superhorizon quanta opens up as the wavelength
nears horizon crossing.}

\item{We discuss the implications of these results for scalar and
tensor perturbations, and establish a connection with previous
calculations of non-gaussian correlations.}
\end{itemize}

The article is organized as follows: In section II we introduce the
models, in section III we present the equations of motion, describe
the approach to obtaining the decay law via the DRG and introduce
the $\Delta$ expansion. We consider the decay of a
scalar field coupled to other scalar fields via a cubic coupling in
pure de Sitter inflation. We study the decay of superhorizon
fluctuations as well as of fluctuations with wave vectors deep inside
the Hubble radius. In section IV we apply the results of section III
to the \emph{self-decay} of the  inflaton quantum fluctuations
during quasi de Sitter slow roll inflation. In section V we discuss
the implications of our results for scalar and tensor metric
perturbations as well as the connection between the quantum decay
processes and the emergence of {\bf non-gaussian}
correlations. Our conclusions and further discussions are contained
in section VI. Two appendices are devoted to the calculation of the self-energy
kernel in the  $\Delta$ expansion for arbitrary wave vector
including the order $\Delta^0$.

\section{The models}

We consider a general interacting scalar field theory with cubic
couplings in a spatially flat Friedmann-Robertson-Walker
cosmological space time with scale factor $a(t)$. The cubic
couplings are the lowest order non-linearities. Our study applies to
 two different scenarios, i) the inflaton  $\phi$ coupled to
 another scalar field $\varphi$, ii) the inflaton field self-coupled
 via a trilinear coupling. We consider the fields to be minimally
 coupled to gravity.

In comoving coordinates the action for case i) is given by
\begin{equation}\label{2fields}
A= \int d^3x \; dt \;  a^3(t) \Bigg\{ \frac{1}{2} \;
{\dot{\phi}^2}-\frac{(\nabla \phi)^2}{2a^2}-\frac{1}{2} \; M^2 \;
\phi^2 + \frac{1}{2} \; {\dot{\varphi}^2}-\frac{(\nabla
\varphi)^2}{2a^2}-\frac{1}{2} \; m^2 \; \varphi^2 - g \;  \phi \,
\varphi^2 +J(t) \; \phi + \mathrm{higher \; nonlinear \; terms}
\Bigg\}
\end{equation}
\noindent  and for the case ii),
\begin{equation}\label{1field}
A= \int d^3x \;  dt \;  a^3(t) \Bigg\{ \frac{1}{2} \;
{\dot{\phi}^2}-\frac{(\nabla \phi)^2}{2a^2}-\frac{1}{2} \; M^2 \;
\phi^2 + \frac{g}{3} \;  \phi^3 +J(t) \; \phi+  \mathrm{higher \;
nonlinear \; terms} \Bigg\}
\end{equation}
\noindent The linear term in $\phi$ is a counterterm that will be
used to cancel the tadpole diagram in the equations of motion. The
higher nonlinear terms do not affect our results but they are
necessary to stabilize the theory.

\begin{figure}[ht!]
\begin{center}
\includegraphics[height=3in,width=6in,keepaspectratio=true]{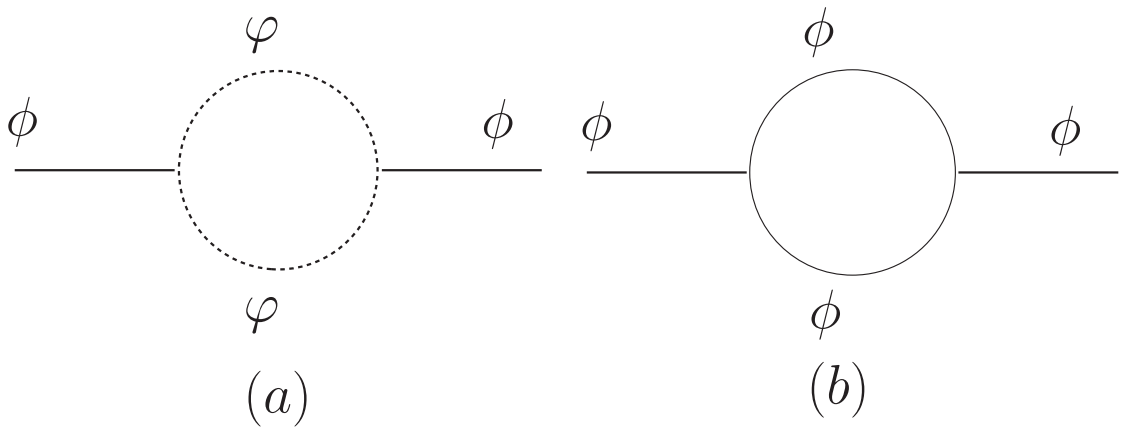}
\caption{Self energies: Fig.(a) depicts the self energy contribution
from a loop of $\varphi$ particles, Fig. (b) depicts the self-energy
from the self-interaction of the inflaton.} \label{fig:selfenergy}
\end{center}
\end{figure}

In order to study the decay of particles associated with a field we
must first obtain the self-energy corrections to the equations of
motion. We  study the decay of inflaton
fluctuations up to one-loop order in the coupling either into other
fields or in self-decay.
The calculation of the self-energy correction up to one loop is
similar in both cases, the extra factor $1/3$ in the
trilinear coupling in Eq. (\ref{1field})
accounts for the combinatorial factor in the
corresponding Feynman diagram. Fig. \ref{fig:selfenergy} shows
the self-energy contributions to the inflaton propagator up to one
loop. Fig. \ref{fig:selfenergy}a displays a loop of $\varphi$
particles and fig. \ref{fig:selfenergy} b displays the one-loop
self energy for a cubic self-interaction which is the lowest order
non-linearity around the classical inflaton (expectation value)
driving inflation.

It is clear from these figures that we only need to obtain the
self-energy in only one of the cases, since one case is
obtained from the other by a simple replacement of the masses
of the particles that run in the loop.
Therefore, we will study the self-energy for the
case of figure fig. \ref{fig:selfenergy} a.

It is convenient to pass to conformal time $\eta$ with $d\eta =
dt/a(t)$ and introduce a conformal rescaling of the fields
\begin{equation}
a(t) \; \phi(\vx,t) = \chi(\vx,\eta)~~;~~
a(t) \; \varphi(\vx,t)=\delta(\vx,\eta)
\end{equation}
The action Eq. (\ref{2fields}) (after discarding surface terms that
do not affect the equations of motion) reads:
\be\label{confoaction}
A\Big[\chi,\delta\Big]= \int d^3x \;  d\eta \;
\Bigg\{ \frac{{\chi'}^2}{2}-\frac{(\nabla
\chi)^2}{2}-\frac{\mathcal{M}^2_{\chi}(\eta)}{2} \; \chi^2 +
\frac{{\delta'}^2}{2}-\frac{(\nabla
\delta)^2}{2}-\frac{\mathcal{M}^2_{\delta}(\eta)}{2} \; \delta^2 -g \;
C(\eta) \;  \chi \; \delta^2 +C^3(\eta) \;  J(\eta) \;  \chi\Bigg\}
\ee
\noindent with primes denoting derivatives with respect to conformal
time $\eta, \; C(\eta)= a(t(\eta))$ is the scale factor as a
function of $\eta$ and
\bea \mathcal{M}^2_{\chi}(\eta)  =  M^2 \;
C^2(\eta)-\frac{C''(\eta)}{C(\eta)} \quad ,  \quad
 \mathcal{M}^2_{\delta}(\eta)  =  m^2 \;
C^2(\eta)-\frac{C''(\eta)}{C(\eta)}  \; .\label{massdelta}
\eea
\noindent In this section we focus on a de Sitter
inflationary cosmology, we treat
slow-roll and quasi de Sitter inflation in sec. IV. For
de Sitter space time the scale factor is given by
\be \label{scalefactor}
a(t)= e^{Ht} \quad ,  \quad C(\eta) = -\frac{1}{H\eta} \; ,
\ee
\noindent with $H$ the Hubble constant and the conformal time $\eta$ is
given by
\be
\eta= -\frac{e^{-Ht}}{H} \; ,
\ee
\noindent where $\eta = -\frac{1}{H} $ corresponds to the initial time $t=0$.

The Heisenberg equations of motion for the Fourier field modes of
wave vector $k$ in the free ($g=0$) theory are given by
\bea
\chi''_{\vk}(\eta)+
\Big[k^2-\frac{1}{\eta^2}\Big(\nu^2-\frac{1}{4} \Big)
\Big]\chi_{\vk}(\eta)& = & 0 \label{chimodes}\; , \\
\delta''_{\vk}(\eta)+
\Big[k^2-\frac{1}{\eta^2}\Big(\bnu^2-\frac{1}{4} \Big)
\Big]\delta_{\vk}(\eta)& = & 0 \label{deltamodes} \eea \noindent
where \be \label{nus} \nu^2  =  \frac{9}{4}- \frac{M^2}{H^2} \quad
, \quad  \bnu^2  = \frac{9}{4}-\frac{m^2}{H^2}  \; . \ee The
Heisenberg free field operators can be expanded in terms of the
linearly independent solutions of the mode equation \be
S''_{\nu}(k;\eta)+ \Big[k^2-\frac{1}{\eta^2}\Big(\nu^2-\frac{1}{4}
\Big) \Big]S_{\nu}(k;\eta) =  0\, , \label{modeeqn} \ee \noindent
for $\nu,\bnu$ respectively. We choose the usual Bunch-Davies
initial conditions for the mode functions, namely the usual plane
waves for wavelengths deep inside the Hubble radius $|k \;
\eta|\gg 1$. The mode functions $S_{\nu}(q,\eta)$ associated with
the Bunch-Davies vacuum are given by \be\label{Snu}
S_{\bnu}(k,\eta)= \frac{1}{2}\; i^{-\nu-\frac{1}{2}} \sqrt{\pi
\eta} \; H^{(2)}_\nu(k\eta) \; . \ee For wavelengths much smaller
than the Hubble radius, ( $|k\eta| \gg 1$) the mode functions with
Bunch-Davies vacuum initial condition behave as plane waves in
Minkowski space time, namely
 \be
S_{\nu}(k;\eta) \buildrel{|k\eta|\gg 1}\over=
\frac{1}{\sqrt{2k}}\,e^{-ik\eta}\;.
\ee
In particular for $\nu = 3/2$, eq.(\ref{Snu}) becomes
\be\label{S32}
S_{\frac{3}{2}}(k,\eta)= \frac{1}{\sqrt{2k}} \;
e^{-ik\,\eta}\left[1-\frac{i}{k\eta}\right] \; .
\ee
The spatial Fourier transform of the free Heisenberg field operators
$\chi_{\vk}(\eta), \;\delta_{\vk}(\eta)$ are therefore written as
\bea
\chi_{\vk}(\eta) & = &
\alpha_{\vk} \; S_{\nu}(k;\eta)+\alpha^\dagger_{-\vk} \; S^*_{\nu}(k;\eta)
\cr \cr
\delta_{\vk}(\eta) & = &
\beta_{\vk} \; S_{\bnu}(k;\eta)+\beta^\dagger_{-\vk} \; S^*_{\bnu}(k;\eta)
\eea
\noindent where the Heisenberg operators
$\alpha_{\vk};\alpha^\dagger_{\vk}$ and
$\beta_{\vk};\beta^\dagger_{\vk}$ obey the usual canonical
commutation relations.
The Bunch-Davies vacuum state $|0\rangle$ is annihilated by
$\alpha_{\vk},\beta_{\vk}$.

\section{Equations of motion, dynamical renormalization group and decay
laws: the $\Delta$ expansion in de Sitter inflation.}

As described in detail in \cite{desiterI}, in a rapidly expanding
cosmology the notion of decay `rate' requires a careful analysis.
In Minkowski space time the decay rate is obtained from the
transition probability per unit time, or alternatively from the
imaginary part of the space-time Fourier transform of the
self-energy evaluated on the particle's mass shell.  In the
transition probability,  the square of the energy-momentum delta
functions accounts for the space-time volume times an overall
delta function of energy-momentum conservation: the transition
probability divided by this volume is the decay rate. In Minkowski
space-time, the self-energy is a function of the difference of the
space-time coordinates due to translational invariance. Hence, a
space-time Fourier transform is available, from which the
imaginary part is extracted. Energy-momentum conservation is of
paramount importance to define the decay rate in Minkowski
space-time, and to determine the kinematic thresholds for particle
production and decay.

\medskip

In a rapidly expanding cosmology the lack of a global time-like
Killing vector prevents energy momentum conservation, although
energy is covariantly conserved. As a result: i) a new definition
of the decay `rate' that does not rely on energy-momentum
conservation, and a different approach to studying the decay law
is necessary  ii) since energy is no longer conserved, novel
processes are possible which are forbidden in Minkowski
space-time, therefore we expect novel \emph{decay channels} which
are absent in Minkowski space-time.

In ref.\cite{desiterI} the decay of a particle  into massless conformally
coupled particles was studied as a test example to present the main concepts:
the mode functions are the same as in Minkowski space time, this
simplified the calculation of the self-energy kernel, allowed a systematic
study of the reliability of the dynamical renormalization group method
and a direct comparison to the Minkowski limit.
This simple case, however, does not feature several important
aspects of the more relevant situation of the dynamics of quantum fields
which are massless or nearly so but minimally coupled to gravity.
This latter case features infrared divergences which are not present
in the simpler case of conformally coupled massless
fields \cite{prem,IRcosmo}. In this article we study  the
case in which the inflaton is massive and minimally coupled to gravity
 which is precisely the relevant case for studying the {\it decay of
quantum fluctuations} during slow roll inflation.

\medskip

While the main aspects of the dynamical renormalization group
method to study decay  were introduced in \cite{desiterI}, we
briefly highlight here the main aspects relevant to this work.

The method relies on studying the real time relaxation of the
expectation value of a field induced by an external source in
linear response: as the source is switched-off the expectation
value relaxes revealing the decay law of the amplitude. While an
exact solution of the equations of motion is readily available in
Minkowski space time because the self-energy is a function of the
difference of the time coordinates (allowing the use of
Fourier-Laplace transforms), this is , in general, not the case
during inflation. Solving the equation of motion in a perturbative
expansion in the coupling, secular terms emerge, these terms grow
in time when the conformal time $\eta \rightarrow 0$,
 limiting the validity of the perturbative expansion. The
dynamical renormalization group precisely allows a
systematic {\bf resummation of these secular terms} and
the uniform asymptotic expansion provided by the
resummation lead to the identification of the decay law of the amplitude.
Thus, the main steps of the method are the following:

\begin{itemize}
\item{First, obtain the  (retarded) equations of motion for the
expectation value of the field in linear response  after
switching-off the source that induces the expectation value. }

\item{ Second, obtain a perturbative expansion of the solution in terms of the
coupling. Such perturbative solution features \emph{secular} terms,
namely terms that grow in time (when conformal time $\eta
\rightarrow 0$ during inflation) and limit the validity of the
perturbative expansion. }

\item{Third, implement the dynamical renormalization group (DRG)
to provide a systematic resummation of these secular terms.
The solution of the DRG equation gives the decay law of the amplitude of
quantum fluctuations.  }

\end{itemize}

 We implement these steps in the general case described by the action  Eq.
 (\ref{2fields}) which couples two fields: $\phi$ and  $\varphi$ with masses
$M$ and $m$, respectively and a
 cubic coupling $g \phi \,\varphi^2$ in exact de Sitter space-time.
In section IV we apply these results to the case of
 the self-coupling of inflaton fluctuations.

 We focus on obtaining the decay law for the quantum fluctuations of the
field $\phi$ which will be later identified with the inflaton field.
We derive the equation of motion for the expectation value of
the field using the non-equilibrium generating functional which
 involves  forward and backward time evolution, typical
 of a density matrix. Unlike the S-matrix  case (which is an in-out
transition  where only forward time evolution is required),
the time evolution of an expectation value is an initial value problem which
requires an in-in matrix element. Real time
 equations of motion obtained from the non-equilibrium generating
 functional are guaranteed to be retarded.

It is convenient to write the spatial Fourier transform of the conformally
re-scaled field $\chi$ as follows
 \be\label{split}
\chi^{\pm}_{\vk}(\eta) = X_{\vk}(\eta) +
 \sigma^{\pm}_{\vk}(\eta)~~;~~ \langle \chi^{\pm}_{\vk}(\eta)\rangle
 =X_{\vk}(\eta) ~~;~~\langle \sigma^{\pm}_{\vk}(\eta)\rangle =0
 \ee
 \noindent where the superscripts $\pm$ refer to the forward and
 backward time branches in the non-equilibrium generating
 functional respectively. The expectation value is the same for both
 branches since the c-number external source  is the same.
The equation of motion for
 the expectation value $X_{\vk}(\eta)$ is obtained by requiring
 $\langle \sigma^{\pm}_{\vk}(\eta)\rangle =0 $ systematically
 order by order in perturbation theory. This is the basis of the
 tadpole method to obtain the equations of motion, which up to $\mathcal{O}(g^2)$ (one-loop) are given by [see  Eq.(37) in
ref.\cite{desiterI}]

\be \label{eqnofmot}
X''_{\vk}(\eta)+\left[k^2-\frac{\nu^2-\frac{1}{4} }{\eta^2}\right]
X_{\vk}(\eta)+ \frac{2 \; g^2  }{\eta \; H^2 } \;
\int_{\eta_0}^{\eta} \frac{d\eta'}{\eta'} \;
\mathcal{K}_{\bnu}(k;\eta,\eta') \; X_{\vk}(\eta')=0\;, \ee
\noindent where the counterterm $J(\eta)$ in the action
(\ref{confoaction}) has been used to cancel the tadpole term
proportional to $\langle \delta^2(\vec{x},\eta)\rangle$, this is
independent of $X_{\vk}(\eta)$ and acts as a source term in the
equation of motion. $ \nu $ and $ \bnu $ are given by
Eq.(\ref{nus} ).

The retarded one-loop self-energy kernel
$\mathcal{K}_{\bnu}(k;\eta,\eta')$ is given by \cite{desiterI}
\be\label{kernel} \mathcal{K}_{\bnu}(k;\eta,\eta') = 2 \int
\frac{d^3q}{(2\pi)^3} \; \mathrm{Im}\left[ S_{\bnu}(q,\eta)
S^*_{\bnu}(q,\eta')S_{\bnu}(|\vq-\vk|,\eta)S^*_{\bnu}(|\vq-\vk|,\eta')\right]
\ee \noindent  and is depicted in fig. \ref{fig:selfenergy}-a. The
mode functions $S_{\bnu}(q,\eta)$ are given by Eq. (\ref{Snu}). We
consider  $ M^2/H^2 \ll 1$ and $ m^2/H^2 \ll 1$ which for the
inflaton case to be studied below corresponds to the slow-roll
approximation, and define \be\label{delta} \Delta \equiv
\frac{3}{2}-\bnu \quad ,  \quad \Delta = \frac{1}{3}
\frac{m^2}{H^2}+ \mathcal{O}\left(\frac{m^4}{H^4}\right) \ee hence
$ \Delta \ll 1$. This small parameter  $\Delta$ will  be related
with the slow-roll parameters and plays an important role in
regulating the infrared behavior in the self-energy. Anticipating
a renormalization of the inflaton mass we write \be\label{massren}
M^2= M^2_R+ g^2 \; \delta M^2_1 + \mathcal{O}(g^4) \Rightarrow
\nu^2 = \nu^2_R - g^2  \; \frac{\delta M^2_1}{H^2}
+\mathcal{O}(g^4)\;, \ee \noindent with $\nu^2_R = 9/4 -
M^2_R/H^2$. The equation of motion up to order $g^2$ becomes \be
\label{eqnofmotren}
X''_{\vk}(\eta)+\left[k^2-\frac{\nu^2_R-\frac{1}{4}
}{\eta^2}\right] X_{\vk}(\eta)+ g^2 \frac{\delta
M^2_1}{H^2\,\eta^2}X_{\vk}(\eta) +\frac{2 \, g^2 }{\eta\;H^2}\;
\int_{\eta_0}^{\eta} \frac{d\eta'}{\eta'} \;
\mathcal{K}_{\bnu}(k;\eta,\eta') \; X_{\vk}(\eta')=0\;, \ee In
what follows we suppress the subscript $R$ to avoid cluttering the
notation, therefore  $\nu$ and the mass must be understood as the
renormalized ones. A perturbative solution of Eq.(\ref{eqnofmot})
is obtained by writing \be\label{pertsol} X_{\vk}(\eta)=
X_{0,\vk}(\eta)+g^2 \; X_{1,\vk}(\eta)+\mathcal{O}(g^4) \ee
\noindent and comparing powers of $g$ leads to a hierarchy of
coupled equations: up to second order in $g$, (one loop order),
they are
\begin{eqnarray}
&&X''_{0,\vk}(\eta)+\left[k^2-\frac{1}{\eta^2}\Big(\nu^2-
\frac{1}{4} \Big) \right]X_{0,\vk}(\eta)
 =  0  \; ,\label{X0}\\
&&X''_{1,\vk}(\eta)+\left[k^2-\frac{1}{\eta^2}\Big(\nu^2-
\frac{1}{4} \Big) \right] X_{1,\vk}(\eta) = \mathcal{R}_1(k,\eta)
\label{X1}\;,
\end{eqnarray}
\noindent where the inhomogeneity is given by
\be\label{source}
 \mathcal{R}_1(k,\eta) = -\frac{\delta
M^2_1}{H^2\,\eta^2} \; X_{0,\vk}(\eta)  -\frac{ 2
}{H^2\;\eta}\int_{\eta_0}^{\eta} \frac{d\eta'}{\eta'} \;
\mathcal{K}_{\bnu}(k;\eta,\eta') \; X_{0,\vk}(\eta') \; . \ee The
mass counterterm $\delta M^2_1$ is fixed by requiring that it
cancels the term proportional to $X_{0,\vk}(\eta)/\eta^2$ arising
from the integral in Eq.(\ref{source}). The first order correction
to the solution is given by 
\be  \label{forsol} X_{1,\vk}(\eta)=
\int_{\eta_0}^{0} d\eta' \; \mathcal{G}_{\nu}(k;\eta,\eta') \;
\mathcal{R}_{1}(k,\eta') \; . 
\ee 
\noindent where
$\mathcal{G}(k;\eta,\eta')$ is the retarded Green's function
obeying  \be\label{GF}
\left[\frac{d^2}{d\eta^2}+k^2-\frac{1}{\eta^2}\Big(\nu^2-
\frac{1}{4} \Big) \right]\mathcal{G}_{\nu}(k;\eta,\eta')=
\delta(\eta-\eta')~~,~~ \mathcal{G}_{\nu}(k;\eta,\eta')=0 \quad
\mathrm{for}\quad \eta'>\eta \; . \ee To compute $X_{1,\vk}(\eta)$
we first need the kernel $\mathcal{K}_{\bnu}(k;\eta,\eta')$. As it
will become clear in the following, this kernel involves
(logarithmic) infrared divergences for $\bnu =3/2$, but it is an
\emph{analytic} function of $\Delta$ that features simple poles at
$\Delta = 0$. Since in slow roll $\Delta \ll 1$ we will use the
parameter $\Delta$ as a regulator much in the same manner as the
$\varepsilon$ expansion in dimensional regularization.   We will
therefore compute the kernel $\mathcal{K}_{\bnu}(k;\eta,\eta')$ at
\emph{leading order} in $\Delta$ by extracting the poles and the
logarithmic terms;
 terms proportional to powers of $\Delta$
give subleading contributions. This is akin to the minimal
subtraction in dimensional regularization.

\subsection{Secular terms, DRG and decay law:}

Let $g_{\nu}(k,\eta)$ and $f_{\nu}(k,\eta)$  be two independent
solutions of the zeroth order equation (\ref{X0}), the most general
solution is
\be
X_{0,\vk}(\eta)= A_{\vk} \; g_{\nu}(k,\eta)+ B_{\vk}\;  f_{\nu}(k,\eta)
\ee
where $  A_{\vk} $ and $  B_{\vk} $ are arbitrary constants.
The linear structure of the perturbative series indicates that the
perturbative solution of the equation of motion has the form
\be
X_{\vk}(\eta) = A_{\vk} \;  g_{\nu}(k,\eta)[1+g^2\;
F_1(k,\eta)+\mathcal{O}(g^4)]+B_{\vk}\;
f_{\nu}(k,\eta)[1+g^2\; H_1(k,\eta)+\mathcal{O}(g^4)]
\ee
The functions $F_1(k,\eta); \; H_1(k,\eta)$ are determined by the first
order solution Eq.(\ref{forsol}) and they feature secular terms,
namely divergent terms in the limit $\eta \rightarrow 0$.
Therefore we write,
\be
F_{1}(k;\eta)= F_{1,s}(k;\eta)+F_{1,f}(k;\eta)~~;~~
H_{1}(k;\eta)= H_{1,s}(k;\eta)+H_{1,f}(k;\eta)\label{extract}
\ee
\noindent where $F_{1,s}(k;\eta),H_{1,s}(k;\eta)$ are secular terms,
whereas $F_{1,f}(k;\eta),H_{1,f}(k;\eta)$ remain bounded as
functions of conformal time. The dynamical renormalization group
absorbs the secular terms into a renormalization of the amplitudes
$A_{\vk},B_{\vk}$ at a given time scale $\widetilde{\eta}$,
(wave-function renormalization) \cite{desiterI,DRG}, namely
\bea
A_{\vk} & = &
A_{\vk}(\widetilde{\eta}) \; Z^A_{\vk}(\widetilde{\eta}) \quad ,\quad
Z^A_{\vk}(\widetilde{\eta})= 1+ g^2  \; z^A_{1,\vk}(\widetilde{\eta})+
\mathcal{O}(g^4) \label{ZA}\\
B_{\vk} & = &  B_{\vk}(\widetilde{\eta}) \; Z^B_{\vk}(\widetilde{\eta})
\quad ,\quad  Z^B_{\vk}(\widetilde{\eta})= 1+ g^2 \;
z^B_{1,\vk}(\widetilde{\eta})+ \mathcal{O}(g^4) \label{ZB}
\eea
The coefficients
$z^A_{1,\vk}(\widetilde{\eta}), \; z^B_{1,\vk}(\widetilde{\eta})$ are
chosen to cancel the secular terms in the perturbative solution at
$\eta=\widetilde{\eta}$ order by order in the perturbative expansion.
Since the scale $\widetilde{\eta}$ is arbitrary and the perturbative
solution does not depend on this scale, the renormalized amplitudes
$A_{\vk}(\widetilde{\eta}), \; B_{\vk}(\widetilde{\eta})$ obey the
following dynamical renormalization group equation to lowest order
(Eqs.(55)-(56) of ref.\cite{desiterI,DRG}),
\bea
\frac{\partial A_{\vk}(\widetilde{\eta})}{\partial
\widetilde{\eta}}-g^2 \;  A_{\vk}(\widetilde{\eta}) \; \frac{\partial F_{1,s}
(k;\widetilde{\eta})}{\partial \widetilde{\eta}}+\mathcal{O}(g^4)  &
= & 0 \\
\frac{\partial B_{\vk}(\widetilde{\eta})}{\partial
\widetilde{\eta}}-g^2 \; B_{\vk}(\widetilde{\eta}) \; \frac{\partial H_{1,s}
(k;\widetilde{\eta})}{\partial \widetilde{\eta}}+\mathcal{O}(g^4)  &= & 0
\eea
The solution of these DRG equations is given by
\bea\label{drgamps}
A_{\vk}(\widetilde{\eta})=
A_{\vk}(\widetilde{\eta}_0) \; e^{g^2\left[F_{1,s}
(k;\widetilde{\eta})-F_{1,s} (k;\widetilde{\eta}_0) \right]+\mathcal{O}(g^4)}
\quad , \quad
B_{\vk}(\widetilde{\eta})= B_{\vk}(\widetilde{\eta}_0)\; e^{g^2\left[H_{1,s}
(k;\widetilde{\eta})-H_{1,s} (k;\widetilde{\eta}_0) \right]+\mathcal{O}(g^4)}
\eea
Setting $\widetilde{\eta}=\eta$
we  obtain the renormalization group improved solution,
\bea
X_{\vk}(\eta) & = & A_{\vk}({\eta}) \;
\,g_{\nu}(k;\eta)\left[1+g^2 \;
F_{1,f}(k,\eta)+\mathcal{O}(g^4)\right]+
B_{\vk}({\eta}) \; f_{\nu}(k;\eta)\left[1+g^2 \;
H_{1,f}(k,\eta)+\mathcal{O}(g^4)\right]\label{RGsol}\\
A_{\vk}({\eta})& = &  A_{\vk}({\eta}_0)~e^{g^2\left[F_{1,s}
(k;{\eta})-F_{1,s} (k;{\eta}_0) \right]}\label{Aamp} \quad , \quad
 B_{\vk}({\eta})  =
B_{\vk}({\eta}_0)~e^{g^2\left[H_{1,s} (k;{\eta})-H_{1,s}
(k;{\eta}_0) \right]} \; .
\eea
\noindent The terms in the brackets in Eq. (\ref{RGsol}) are truly
perturbatively small at all conformal times. The real part of the
exponential factors in the complex amplitudes
Eq. (\ref{Aamp})  determine the decay law  of the amplitude (or growth law in
the case of instabilities).

\subsection{The $\Delta$ expansion:}

When the  inflaton decays into minimally coupled \emph{massless} particles,
infrared divergences in the self-energy kernel are present \cite{prem}. These
divergences are a hallmark of minimally coupled massless particles
namely $\bnu=\nu=3/2$  in the intermediate state, and are
similar to those found for gravitons in de Sitter space-time\cite{IRcosmo}.
We are instead considering the case in which {\it both}
the inflaton and the decay products are \emph{massive} with masses
$M$ and $m$ respectively and $(M^2/H^2,m^2/H^2) \ll 1$. The mass
of the particles in the loop cutoffs the infrared divergences here,
(since $m^2/H^2 \neq 0$ then $\bnu \neq 3/2 $ and $ \Delta\neq 0$ [Eq.
(\ref{delta})]). As it will become clear in the explicit calculations
below, the infrared divergences in the self-energy kernel
manifest as simple poles at $\Delta=0$.
Thus, $0<\Delta \ll 1 $ emerges as an \emph{infrared
regulator} akin to the dimensional regularization parameter
$\varepsilon = 4-D$ in the loop expansion in D-dimensional
Minkowski space-time.

Since $\Delta \ll 1$ for slow roll inflation,
we  compute the self-energy kernel in an
expansion in $\Delta$ keeping the poles at $\Delta=0$ and the leading
logarithms in $\eta$ just like the $\varepsilon$ expansion in dimensional
regularization. (The leading $ \log \eta $ terms are the remnant of the
infrared divergence regulated by $\Delta$ just like in
the $\varepsilon$ expansion of critical phenomena).

The details of the calculation of
$\mathcal{K}_{\bnu}(k,\eta,\eta')$ in the $\Delta$-expansion for
arbitrary $k$ is presented in the appendix and the result for the
kernel is given by equations (\ref{finK}) and (\ref{finKlead}).
We now have all the elements necessary to study the decay law.

\subsection{Superhorizon modes: $k=0$}

We begin by the superhorizon modes and take $k=0$.
The general solution of the unperturbed mode Eq.(\ref{X0})
and the retarded Green's function $\mathcal{G}_{\nu}(0,\eta,\eta')$
Eq. (\ref{GF})  for $k=0$ are given by
\bea \label{GF0}
X_{0,\vec{0}}(\eta) &=& A\;(-\eta)^{\beta_+}+B\;(-\eta)^{\beta_-} \; ;
\; \beta_{\pm} = \frac{1}{2}\pm \nu. \label{X00} \\
\mathcal{G}_{\nu}(0,\eta,\eta')&=&
\frac{1}{2\nu}\left[(-\eta)^{\beta_+}\;(-\eta')^{\beta_-}-(-\eta)^{\beta_-}\;
(-\eta')^{\beta_+}\right]\Theta(\eta-\eta')
\eea
We compute  the kernel $\mathcal{K}_{\bnu}(0,\eta,\eta')$ in Appendix A
highlighting the most relevant physical processes and displaying the origin
of infrared divergences as poles at $\Delta=0$ and the leading logarithms
in $\eta$.  $\mathcal{K}_{\bnu}(0,\eta,\eta')$ can also be obtained
in the limit $k\rightarrow 0$ of the $ k \neq 0 $ kernel treated in
Appendix B  [Eq.(\ref{finK})].

For $k=0$ the self-energy kernel is given by
\be\label{K0}
\mathcal{K}_{\bnu}(0;\eta,\eta') = \frac{1}{\pi^2}
\int_0^{\infty} q^2  \; dq ~
\mathrm{Im}\left\{\left[S_{\bnu}(q,\eta)S^*_{\bnu}(q,\eta')
\right]^2\right\} \quad ,  \quad \bnu=\frac{3}{2}-\Delta \; .
\ee
The infrared divergences at $ \Delta = \frac{3}{2} - \bnu = 0 $
arise from the small momenta behavior of the integrand in Eq.(\ref{K0}).
Keeping $ \Delta $ small but nonzero, we find for the kernel (see appendix A)
\be\label{Kbarnu}
\mathcal{K}_{\bnu}(0;\eta,\eta')=\mathcal{K}_{\frac{1}{2}}(0;\eta,\eta')+
\frac{1}{6\pi^2} \Bigg\{   \left[\frac{1}{2\Delta}+\frac23 \right]
\left(\frac{\eta'}{\eta^{2}}-\frac{\eta}{\eta^{'2}}\right)
- \frac{\eta'}{\eta^2}\ln\left(\frac{\eta'}{\eta}  \right)+
\left(\frac{\eta}{\eta^{'2}}- \frac{\eta'}{\eta^2} \right)
\ln\left[1-\frac{\eta}{\eta'} \right]
+ \frac{1}{\eta'}-\frac{1}{\eta} \Bigg\}
\ee
\noindent where
\be\label{K12}
\mathcal{K}_{\frac{1}{2}}(0;\eta,\eta')=-\frac{1}{8\pi^2} \;
\mathcal{P}\left( \frac{1}{\eta-\eta'}\right) = -\frac{1}{8\pi^2} \;
\frac{\eta-\eta'}{\left(\eta-\eta'\right)^2 + (\epsilon \;
\eta')^2}=  -\frac{1}{8\pi^2} \;
\frac12\left[\frac{1}{\eta-\eta'+i\epsilon \; \eta'}+
\frac{1}{\eta-\eta'-i\epsilon \; \eta'}\right] ~~;~~ \epsilon
\rightarrow 0\; .
\ee
This prescription for the principal part regulates
the short distance divergence in the operator product expansion with
a dimensionless infinitesimal  $\epsilon$ independent of
time.  This choice of regularization is consistent with the
short-distance singularities of the operator product expansion in
Minkowski space-time, and  leads to a time-independent mass
renormalization (see also ref. \cite{desiterI}).

The two  terms in the Eq.(\ref{Kbarnu}), namely
$\mathcal{K}_{\frac{1}{2}}(0;\eta,\eta')$ and the term in
braces have very different origin. $\mathcal{K}_{\frac{1}{2}}(0;\eta,\eta')$
accounts for
the large loop momentum contribution $q\eta,q\eta' \gg 1$ where
the behavior of the mode functions is the same as for
conformally coupled massless fields, in particular, the short
distance (ultraviolet) divergence is present only in this term. The
terms in the braces account for the strong infrared behavior
of superhorizon wavelengths reflected by  the pole at $\Delta=0$ and the
logarithms in $ \eta $. This calculation exhibits clearly the origin of the
different contributions.

\begin{figure}[ht!]
\begin{center}
\includegraphics[height=2.5in,width=3in,keepaspectratio=true]{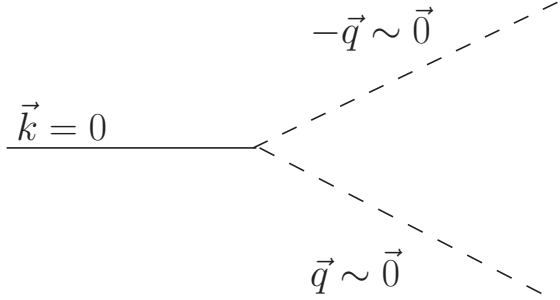}
\caption{Decay of the field $\phi$ (solid line) for $k=0$ into
superhorizon modes of the field $\varphi$}(dashed lines)
\label{fig:fivfi}
\end{center}
\end{figure}

It remains to integrate over $\eta'$ in the second term in Eq.
(\ref{source}) with the kernel given by Eq.(\ref{Kbarnu}). The
integral involving $\mathcal{K}_{\frac{1}{2}}(0;\eta,\eta')$ in
Eq.(\ref{Kbarnu}) was given in ref.\cite{desiterI}. The integrals
over $\eta'$ for the second term (between braces) in
Eq.(\ref{Kbarnu}) can be done easily by expanding the
$\ln[1-\eta/\eta']$ in a power series in $\eta/\eta'$ and
integrating term by term. The result of the integral over $\eta'$
in Eq. (\ref{source}) is of the form \be\label{integ} \frac{ 2
}{H^2\;\eta}\int_{\eta_0}^{\eta} \frac{d\eta'}{\eta'} \;
\mathcal{K}_{\bnu}(0;\eta,\eta') \; X_{0,0}(\eta')= A \;
(-\eta)^{\beta_+} \; \frac{\alpha_+}{\eta^2}+B \;  (-\eta)
^{\beta_-} \; \frac{\alpha_-}{\eta^2}+F[\eta,\eta_0] \ee \noindent
where $F[\eta,\eta_0]$ refers to the contribution of the lower
integration limit and does not produce secular terms in
$X_{\vec{0}}(\eta)$.

Integrating over $ \eta' $ in Eq.(\ref{integ}) yields,
\be\label{alfa}
\alpha_+= \frac{1}{(2 \, \pi \, H)^2}\left[ \ln \epsilon + \gamma +
\psi(\frac12-\nu)\right] + \frac{1}{3(\pi \, H)^2}\left[
\frac{1}{\nu^2-\frac94}\left(\frac{3}{2 \,\Delta} + 2 -3 \,\gamma\right)
+\frac{1}{\nu^2-\frac14}+\frac{1}{(\nu+\frac32)^2}+
\frac{\psi(\frac52 - \nu)}{\frac{3}{2}-\nu}
+\frac{\psi(-\frac12 - \nu)}{\frac{3}{2}+\nu}\right]  \; .
\ee
where we used Eq.(\ref{Kbarnu}), $\gamma$ is the Euler-Mascheroni constant,
and the contribution from
$\mathcal{K}_{\frac{1}{2}}(0,\eta,\eta')$ \cite{desiterI}.
$ \alpha_- $ follows from $ \alpha_+ $ by changing $ \nu \to -\nu $ while
$ \bnu $ is unchanged.
Introducing the symmetric and antisymmetric combinations,
\be  \label{symJ}
\alpha_s  =
\frac{1}{2} \left[\alpha_++\alpha_-\right] \quad ,  \quad
\alpha_a  =   \frac{1}{2} \left[ \alpha_+-\alpha_-\right] \; ,
\ee
\noindent the symmetric term,
\be\label{sime}
\left[ A \;  (-\eta)^{\beta_+}+B \; (-\eta)^{\beta_-} \right] \;
\frac{\alpha_s}{\eta^2}=
\frac{X_{0,\vec{0}}(\eta)}{\eta^2} \; \alpha_s
\ee
\noindent is identified with a contribution to mass renormalization
and is cancelled by the counterterm $\delta M^2_1$ including
the logarithmic ultraviolet divergence $\ln \epsilon $
[$\epsilon$ the short distance regulator Eq. (\ref{K12})].
Eqs.(\ref{alfa}) and (\ref{symJ}) yield,
\be \label{Jasy}
\alpha_a = \frac{1}{8\pi\,H^2} \tan[\pi\,\nu]\left[1+
\frac{4}{\frac{9}{4}-\nu^2}\right] \; .
\ee
The unit term  in the bracket arises from the contribution of
$\mathcal{K}_{\frac{1}{2}}(0,\eta,\eta')$.
 After cancelling the term given by Eq. (\ref{sime})
with a proper choice of the mass counterterm,  and
taking into account that $F[\eta,\eta_0]$ does not contribute to
the secular terms, we find the
 solution of the equation of motion up to $\mathcal{O}(g^2)$
 \be\label{solg2}
X_{\vec{0}}(\eta)=X_{0,\vec{0}}(\eta)\left[1+\Gamma \;
 \ln\frac{\eta}{\eta_0}+\mathrm{non-secular~terms}\right]
\ee
 \noindent with
 \be\label{Gamma_2}
\Gamma = \frac{g^2}{16\pi\,H^2\,\nu}\tan[\pi\,\nu] \left[1+
\frac{4}{\frac{9}{4}-\nu^2}\right]
= \frac{g^2}{16\pi\,H^2\,\nu}
\tan[\pi\,\nu]\left[1+ \frac{4\,H^2}{M^2}\right] ~~, ~~
\nu = \sqrt{\frac{9}{4}-\frac{M^2}{H^2}} \; .
\ee
The dynamical renormalization group resummation exponentiates the secular terms
in Eq.(\ref{solg2}) [see Eqs.(\ref{RGsol})-(\ref{Aamp})]
and leads to the  improved solution,
\be
X_{\vec{0}}(\eta)=
\left[\frac{\eta}{\eta_0}\right]^{\Gamma}\Bigg\{A(\eta_0)
\;  (-\eta)^{\beta_+} [1+\mathcal{O}(g^2)]+B({\eta}_0) \;
 (-\eta)^{\beta_-}[1+ \mathcal{O}(g^2)]\Bigg\}\;.\label{resumX0ren}
\ee
The first term inside the square bracket in Eq. (\ref{Gamma_2})
(namely the unit term) corresponds to the case in which the
inflaton decays into massless particles conformally coupled to
gravity \cite{desiterI},\cite{prem}.


The calculation leading to eq.(\ref{Gamma_2}) is valid for $ \bnu \to \frac32 $
(namely, $ m \ll H $) and we keep $ \nu $ as well as $M$ arbitrary.
We can analytically continue the formula (\ref{Gamma_2})
to $ H < M $ and then take the $ m \ll H \ll M $ limit. In this limit
$\Gamma$ becomes the decay rate of a particle with mass $M$ into massless
particles in Minkowski space-time:
$$
\lim_{m \ll H \to 0} H \; \Gamma = \Gamma_{Mink} =  g^2/(16 \pi M) \quad ,
$$
as it must be.


\subsection{Modes inside the horizon during inflation: $|k\eta| \gg 1$}

\begin{figure}[ht!]
\begin{center}
\includegraphics[height=3in,width=4in,keepaspectratio=true]{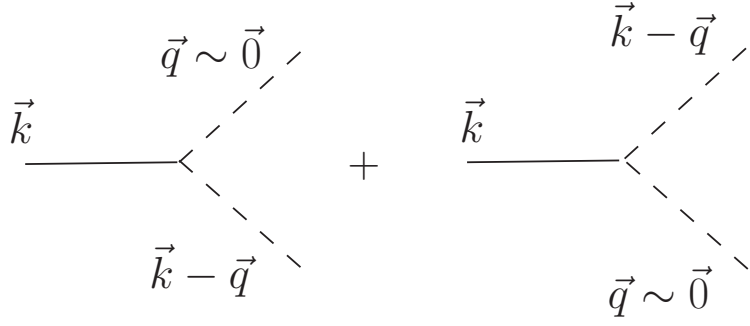}
\caption{Infrared contributions to $\phi \rightarrow \varphi
\varphi$. The external particle has a wavelength deep inside the
horizon but one of the intermediate lines has superhorizon
wavelengths. This process is identified as \emph{bremsstrahlung}
radiation of superhorizon quanta. } \label{fig:fidecayK}
\end{center}
\end{figure}

The kernel $\mathcal{K}_{\bnu}(k;\eta,\eta')$ for arbitrary $k$
has been computed in the appendix in leading order in the $\Delta$
expansion and up to leading logarithms. It is  given by equations
(\ref{finK}) and (\ref{finKlead}). Obtaining the perturbative
solution and extracting the secular terms leading to the decay law
for arbitrary $k$ is an extremely difficult task which requires a
full numerical study. However, explicit expressions can be derived
for wavelengths deep inside the Hubble radius all throughout
inflation, namely $|k\eta|,|k\eta'| >> 1$. In the appendix we show
that in the short wavelength limit the kernel simplifies to the
following expression [Eq. (\ref{kernelhik})] \bea
\label{kernelhik2} \mathcal{K}_{\bnu}(k,\eta,\eta')   =
\mathcal{K}_{\frac{1}{2}}(k,\eta,\eta')   & - &  \frac{1}{4\pi^2
k\eta\eta'} \Bigg\{\sin
k(\eta-\eta')\left[\frac{1}{\Delta}+\mathcal{C}-\ln
k(\eta-\eta')+\ln k^2\eta\eta'
-\mathrm{Ci}[2k(\eta-\eta')]\right]+\nonumber\\ && +\cos
k(\eta-\eta')\left[ \frac{\pi}{2}+\mathrm{Si}[2k(\eta-\eta')]
\right]\Bigg\} \; , \eea \noindent where
$\mathcal{C}=\ln2+\gamma-2$.  Again, the first term in this
expression is the self-energy kernel for conformally coupled
massless fields in the loop, \be
\mathcal{K}_{\frac{1}{2}}(k,\eta,\eta')=-\frac{1}{8\pi^2} \; \cos
k(\eta-\eta') \; \mathcal{P}\left(\frac{1}{\eta-\eta'}\right) \; .
\ee The principal part is defined by Eq. (\ref{K12}). A close
examination of the steps leading to this expression in the
appendix shows that this contribution originates solely from the
high loop momenta $ q\eta,q\eta'\gg 1$ for which the mode
functions coincide with those of massless conformally coupled
fields. The second term within brackets which features the
$1/\Delta$ and the logarithms originate in the process of emission
of superhorizon modes. Two regions in the integral over the  loop
momentum $q$ give rise to these contributions: $q \ll 1/\eta$ and
$|\vk -\vq| \ll 1/\eta$, corresponding to the case when either
line in the loop transfers very small momentum (superhorizon
modes). These processes can be described as \emph{bremsstrahlung
radiation} of superhorizon quanta and are depicted in fig.
\ref{fig:fidecayK}.

This process is analogous to the generation of Hawking
radiation from black holes.  In the case of Hawking radiation, a pair is
created from the vacuum, a particle falling inside the horizon and the other
one being emitted outside. In the present case a particle inside the Hubble
radius decays into a pair: a particle goes outside the Hubble radius and
the other inside. Analysis of the phase space integration carried out
in the appendix reveals that the emitted superhorizon quanta are
almost collinear with the (large) external momentum $k$.

In summary: the processes which yield the leading infrared behavior
responsible for the term $1/\Delta$ and the leading
logarithm  in the kernel Eq. (\ref{kernelhik2}) correspond to
{\it collinear bremsstrahlung radiation of superhorizon quanta}.

In the limit $|k\eta| >>1$ the zeroth order equation of motion is
\be\label{zeror}
X''_{0,\vk}(\eta)+ k^2 X_{0,\vk}(\eta) = 0
\ee
\noindent whose solutions are simple plane waves describing short
wavelength  modes deep inside the Hubble radius,
\be\label{zeroorhik}
X_{0,\vk}(\eta)=
A_{\vk} \; e^{-ik\eta}+B_{\vk} \; e^{ik\eta}
\ee
The equation of motion for the first order perturbation is given by
\be\label{firstor}
X''_{1,\vk}(\eta)+ k^2 X_{1,\vk}(\eta) =
\mathcal{R}_1(k,\eta),
\ee
\noindent where the inhomogeneity is given by Eq. (\ref{source})
with $X_{0,\vk}(\eta)$ given by Eq.(\ref{zeroorhik}).
The calculation of $\mathcal{R}_1(k,\eta)$ for
general $k$ is very complicated, but the leading order terms in the
limit $|k\eta| \gg 1$ can be extracted systematically. There are two
distinct contributions : i) the first term in Eq.(\ref{kernelhik2}),
namely the short wavelength modes which yield the kernel
of conformally massless fields
$\mathcal{K}_{\frac{1}{2}}(k,\eta,\eta')$,  and ii) the superhorizon modes
yielding the second term in Eq.(\ref{kernelhik2}) with the $1/\Delta$
and the leading logarithms.

The first term contains a short distance divergence proportional to
$X_{0,\vk}(\eta) \; [\ln\epsilon]/\eta^2$ where $\epsilon$ is defined
in Eq. (\ref{K12}). This term is canceled by the proper
choice of the mass counterterm in the inhomogeneity Eq. (\ref{source}).
The second term does not yield a mass renormalization to leading
order in $\Delta$. After a proper choice of the mass renormalization
counterterm, we find
\bea
\label{R}\mathcal{R}_1(k,\eta)= && \frac{i}{8\pi^2 \eta^2 H^2}
\Bigg\{ A_{\vk} \;e^{-ik\eta} \left[\frac{\pi}{2}-i \; \ln
\frac{\eta}{\eta_0}  \right] - B_{\vk} \; e^{ik\eta}
\left[\frac{\pi}{2}+i \;  \ln
\frac{\eta}{\eta_0}  \right] \Bigg\}\nonumber\\
&&-\frac{i}{4\pi^2kH^2\eta^3} \Bigg\{  A_{\vk}\; e^{-ik\eta}
\left[\frac{1}{\Delta}+\ln(-k\eta)-i \; \frac{\pi}{2} \right]-
B_{\vk} \; e^{ik\eta}
\left[\frac{1}{\Delta}+\ln(-k\eta)+i \; \frac{\pi}{2} \right]\Bigg\}
\eea
\noindent where we have displayed separately the contributions from
the high momentum modes yielding the first term and arising from
$\mathcal{K}_{\frac{1}{2}}(k,\eta,\eta')$, and those
from the superhorizon modes yielding the second term which feature
the hallmark $1/\Delta$ and logarithms. We have kept the leading
order terms in the real and imaginary parts inside the brackets
neglecting terms suppressed by higher powers of
$1/|k\eta|<<1$. A noteworthy feature of the contribution of the
superhorizon modes is the extra factor $1/k\eta$. The origin of the
extra factor $1/k$ can be traced from the phase space angular
integration Eq. (\ref{ang}).  The integrals  yielding
$\mathcal{K}_{\frac{1}{2}}(k;\eta,\eta')$ are dominated by momenta $q \geq k$
which compensate the factor $1/k$. The integration over the
superhorizon modes cannot compensate the $1/k$ for  large $k$
corresponding to modes well within the horizon. Hence, the extra
factor $1/k$ is a consequence of the small phase space available for
the coupling between high and small momentum modes. For dimensional
reasons this extra factor  $k$ appears with an extra factor
$\eta$ which is the only other scale in the integrals. In summary:
the contribution of the superhorizon modes yields a strong infrared
behavior which is regulated by $\Delta$ and is suppressed by phase
space by an extra power of $1/|k\eta| = H C(\eta)/k \ll 1$ in the
limit $|k\eta| \gg 1$, (i. e., for wavelengths much smaller
than the Hubble radius during inflation).

The solution of the equation of motion (\ref{firstor}) is found by
using the retarded Green's function in the short wavelength limit,
which is given by
\be
\mathcal{G}_{\nu}(k;\eta,\eta') = \frac{1}{k} \; \sin k(\eta-\eta') \;
\Theta(\eta-\eta') \; .\label{GFhik}
\ee
\noindent Therefore, the first order correction becomes
\be\label{sol}
X_{1,\vk}(\eta)= \frac{1}{k}\int^{\eta}_{\eta_0} \sin
k(\eta-\eta') \; \mathcal{R}_1(k,\eta') \; d\eta' \;  .
\ee
The final integral with the retarded
Green's function as in Eq. (\ref{sol}) can now be
performed extracting again the leading order terms for
$|k\eta| \gg 1$. We find
\bea\label{secX1} X_{1,\vk}(\eta) = &&-
X_{0,\vk}(\eta)\Bigg\{\frac{1}{32\pi k
H}\left[C(\eta)-C(\eta_0)\right]+\frac{1}{16\pi^2 k^2
}\left[C^2(\eta)\left(\frac{1}{\Delta}+\ln(-k\eta)\right)-C(\eta_0)
\left(\frac{1}{\Delta}+\ln(-k\eta_0)\right)\right]\Bigg\}+\nonumber
\\ && + \; \; \mathrm{non-secular}
\eea
\noindent where $C(\eta)=-1/H\eta$ is the scale factor and we have
omitted  purely imaginary secular terms since
the dynamical renormalization group  exponentiates them [Eq.(\ref{drgamps})]
to a (time dependent) phase. The terms displayed in
Eq.(\ref{secX1}) are truly secular, since $ \log|\eta| $
grows by about $60$ during inflation. From the dynamical
renormalization group resummation Eq.(\ref{drgamps}) we find the
following improved solution of the equations of motion
\bea\label{solu}
X_{\vk}(\eta) & = &
A_{\vk}(\eta)\,e^{-ik\eta}\left[1+\mathcal{O}(g^2)\right]+B_{\vk}(\eta)
\; e^{ik\eta}
\left[1+\mathcal{O}(g^2)\right]\nonumber\\
A_{\vk}(\eta) & = &
A_{\vk}(\eta_0)\,e^{-\left[\Gamma(k,\eta)-\Gamma(k,\eta_0)\right]}
\; e^{i\xi(k,\eta)}\nonumber\\
B_{\vk}(\eta) & = &
B_{\vk}(\eta_0)\,e^{-\left[\Gamma(k,\eta)-\Gamma(k,\eta_0)\right]}\,
e^{-i\xi(k,\eta)}
\eea
\noindent where the real phase $\xi(k,\eta)$ is not relevant for the decay rate,
and the decay law of the amplitudes is given by
\be\label{rate}
\Gamma(|k\eta|\gg 1)  = \frac{g^2}{32\pi H^2} \;
\frac{H}{k_{ph}(\eta)}\left[1+\frac{2H}{\pi
k_{ph}(\eta)}\left(\frac{1}{\Delta}+\ln\frac{k_{ph}(\eta)}{H}\right)\right]
\quad , \quad k_{ph}(\eta)\equiv \frac{k}{C(\eta)}.
\ee
Modes with  $k$ deep within the horizon satisfies
$k_{ph}(\eta)/H \gg 1$. The unit term in
the bracket in Eq. (\ref{rate}) corresponds to the particles in
the loop being massless and conformally coupled to gravity. The
second term which features the factors $1/\Delta$ and the logarithmic term
arise from the emission of superhorizon quanta. We see that the
infrared regularization provided by $\Delta$ yields a  finite result for the
decay law. Furthermore, the factors associated with
the infrared processes are suppressed by an extra power of
$H/k_{ph}(\eta)\ll1$. This suppression is a consequence of the small
phase space available for the coupling between high and small
momentum modes as can be seen directly from Eq. (\ref{ang}).

The contributions to the decay law Eq.(\ref{rate})
from the emission of superhorizon
quanta become larger the closer is the wavelength to horizon
crossing.  For $H/k_{ph}(\eta) \ll 1$ they can even dominate $ \Gamma $ for
sufficiently small $\Delta$.

An important aspect of the decay law, either for modes
inside or outside the Hubble radius, is that there are no
\emph{kinematic thresholds}. This is a consequence of the inflationary
expansion [$(M,m) \ll H$] and the lack of energy conservation. In
Minkowski space-time, energy-momentum conservation leads to kinematic
thresholds, in particular a massive particle cannot decay in its own
quanta. However, in an inflationary cosmology this process is
allowed, namely \emph{a particle can decay into itself}.

\section{Self-decay of inflaton  Quantum fluctuations during
slow-roll inflation. }

Fluctuations of the inflaton in exact de Sitter inflation do not
seed density perturbations: scalar metric perturbations couple
to the  inflaton fluctuations through the time derivative of
the inflaton expectation value. Therefore, the relevant case
for density perturbations  is quasi de Sitter inflation,
in particular slow-roll
inflation \cite{liddle,lidsey}, which serves as the basis
of CMB data analysis \cite{peiris}. Our ultimate goal is to understand how
quantum effects from interactions, such as the decay of
fluctuations, can affect the power spectrum of scalar and tensor
metric perturbations. During slow roll, the scalar field  fluctuations
are sources for scalar metric fluctuations, therefore
quantum effects as studied here can produce  novel
signatures on the power spectrum. While a complete
gauge invariant  description is ultimately required to
treat this issue, we  focus here on the decay of the inflaton quantum
fluctuations, which in longitudinal gauge are directly related to the scalar
metric fluctuations \cite{bran,hu2}.

Therefore, we apply our results above to the  quantum  \emph{self} decay
of the  inflaton fluctuations. We consider only one scalar field,
namely the inflaton
whose Lagrangian density is given by
\be\label{lagrainfla}
\mathcal{L} = \frac{1}{2} \; \partial^{\mu}\Phi\partial_{\mu}\Phi - V[\Phi] \ee
We write
\be\label{zeromode}
\Phi(\vx,t)= \Phi_0(t) + \phi(\vx,t)
\ee
\noindent where $t$ is the cosmic time, $\Phi_0(t)$ is the
expectation value of the inflaton
field which drives the FRW background metric and $\phi(\vx,t)$
describes the inflaton quantum fluctuations.
Expanding around the expectation value, the Lagrangian density for
the quantum fluctuations reads
\be\label{lagfluc}
\delta \mathcal{L}[\phi;\Phi_0]= \frac{1}{2}\left(
\partial^{\mu}\phi \; \partial_{\mu}\phi - V''[\Phi_0] \; \phi^2
\right) +\frac{g}{3} \; \phi^3+\mathrm{higher \; order \; terms} \; ,
\ee
\noindent where the primes applied to the potential $V[\Phi]$
stand for derivatives with respect to the argument (not to
be confused with derivatives with respect to conformal time)
and  we have used the equation
of motion for  $\Phi_0(t)$  which   in cosmic time  is given by
\be\label{eqnfi0}
\ddot{\Phi}_0+3 \; H \; \dot{\Phi}_0(t)+V'[\Phi_0] = 0 \; .
\ee
We have kept the lowest order term in the
non-linearity and defined the (dimensionful) coupling constant
\be\label{coup}
g \equiv \frac12 \; V'''[\Phi_0] \; \; .
\ee
As it is clear from the study in the previous section, the
perturbative treatment of the non-linearity will be reliable
provided $g/H \ll 1$.

In the slow roll approximation the equation of motion simplifies
to \be\label{slowroll} 3 \; H \; \dot{\Phi}_0(t)+V'[\Phi_0] = 0 \;
. \ee The slow-roll parameters relevant to our discussion are the
following (either in terms of $H$ (Hubble)  or $V$
(potential))\cite{liddle,lidsey} \bea &&\epsilon_H = 2 \; M^2_{Pl}
\; \left(\frac{H'}{H} \right)^2 \quad , \quad \epsilon_V  =
\frac{M^2_{Pl}}{2}
\left(\frac{V'[\Phi_0]}{V[\Phi_0]} \right)^2 = \epsilon_H \, \label{epsiv}\\
&& \eta_H =  2 \; M^2_{Pl} \; \frac{H''}{H} \quad , \quad
\eta_V   =   M^2_{Pl}  \; \frac{V''[\Phi_0]}{V[\Phi_0]} = \eta_H + \epsilon_V
\label{etav}\\
&&\xi_H = 4 \; M^2_{Pl} \;\frac{H' \; H'''}{H^2} \quad , \quad
\xi_V  = M^4_{Pl} \; \frac{V'[\Phi_0] \; V'''[\Phi_0]}{V^2[\Phi_0]}=
\xi_H + 3 \; \epsilon_H
\; \eta_H \; .
\eea
 \noindent Here, $ M^2_{Pl} \equiv 1/[8 \, \pi \, G] =
m^2_{Pl}/(8 \, \pi) $ and $ M_{Pl} = 2.4 \; 10^{18}$ GeV. Slow
roll implies $(\epsilon_V,\eta_V ,\xi_V,\epsilon_H,\eta_H ,\xi_H)
\ll 1$, and $\xi_V,\xi_H $ are formally of second order in
 slow-roll.

\begin{figure}[ht!]
\begin{center}
\includegraphics[height=3in,width=3in,keepaspectratio=true]{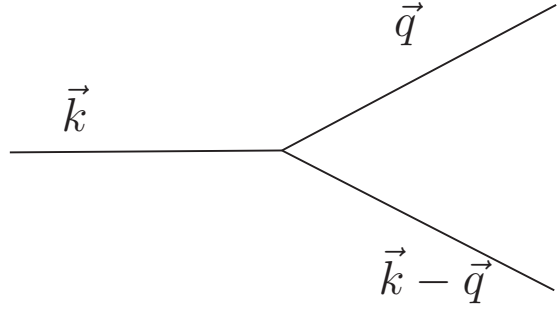}
\caption{Self-decay of quantum fluctuations of the inflaton. All
lines correspond to the field $\phi$, i.e, the quantum fluctuations
of the inflaton. } \label{fig:inflatondecay}
\end{center}
\end{figure}

In terms of the slow roll parameters the Friedmann equation reads
 \be\label{FRW}
H^2 = \frac{V}{3M^2_{Pl}}\left[1+\frac{\epsilon_V}{3}+
\mathcal{O}(\epsilon^2_V,\epsilon_V \; \eta_V)
\right]
 \ee
\noindent and the effective mass of the inflaton quantum fluctuations
is given by
 \be\label{flucmass}
M^2 = V''[\Phi_0] = 3 \; H^2 \;  \eta_V +
\mathcal{O}(\epsilon_V \; \eta_V)  \; .
 \ee
During slow roll inflation the scale factor is quasi de Sitter and
to lowest order in slow roll:
\be\label{quasiDS}
C(\eta)=-\frac{1}{H \; \eta} \; \frac{1}{1-\epsilon_V}= -\frac{1}{H \; \eta}
(1+\epsilon_V) + \mathcal{O}(\epsilon_V^2)
\ee
 We are now in a position to apply the results obtained in the
 previous sections to study the \emph{self} decay of the quantum fluctuations
of the inflaton via the cubic coupling.
 The decay process $\phi\rightarrow \phi \; \phi$  is depicted in fig.
 \ref{fig:inflatondecay}. The self-energy has the same form
 as for an interaction $\phi \;\varphi^2$ analyzed in
 the previous section, the \emph{only} difference is that in the
 self energy we have $\nu$ instead of $ \bnu$.

Within slow roll, the linearized equation of motion for the quantum
 fluctuations of the inflaton  $ \phi $ is precisely given by
\be \label{flucds}
\chi''_{\vk}(\eta)+
\left[k^2 + M^2 \; C^2(\eta) - \frac{C''(\eta)}{C(\eta)}\right]
\chi_{\vk}(\eta) = 0 \; ,
\ee
with the quasi de Sitter scale factor $C(\eta)$ given by Eq.(\ref{quasiDS})
and $M^2$ given by  Eq.(\ref{flucmass}).
Let us compute the expression within brackets in Eq. (\ref{flucds}) to
first order in slow roll. We have in cosmic time:
\be\label{c2c}
\frac{C''(\eta)}{C(\eta)} = a^2(t)\left[ {\dot H} + 2 \; H^2 \right] \; .
\ee
And Eqs. (\ref{slowroll}), (\ref{epsiv}) and (\ref{FRW}) yield,
\be\label{h2h}
\frac{\dot H}{H^2} = - \epsilon_V + \mathcal{O}(\epsilon_V^2) \; .
\ee
Eqs.(\ref{quasiDS}), (\ref{c2c})  and (\ref{h2h}) thus imply,
\be\label{c2c2}
\frac{C''(\eta)}{C(\eta)} = \frac{1}{\eta^2}\left[2 + 3 \; \epsilon_V
+ \mathrm{higher~orders~in~slow~roll}\right]
\ee
and
\be
 M^2 \; C^2(\eta) - \frac{C''(\eta)}{C(\eta)}= -
\frac{1}{\eta^2}\left[2 + 3 \; \left(\epsilon_V- \eta_V\right)
+ \mathrm{higher~orders~in~slow~roll}\right]
\ee
where we also used Eq. (\ref{flucmass}). In summary, during slow roll the
quantum fluctuations of the inflaton behave in a similar way as in pure de
Sitter space-time [sec. II and II] but with the index $\nu$ of the mode
 functions given by
\be \label{nusr}
\nu=\bnu=\frac{3}{2}+ \epsilon_V-\eta_V +
\mathrm{higher~orders~in~slow~roll}\; ,
\ee
where
$$
 M^2 \; C^2(\eta) - \frac{C''(\eta)}{C(\eta)}\equiv -
\frac{\nu^2-\frac14}{\eta^2}+
\mathrm{higher~orders~in~slow~roll} \; .
$$
\noindent  In this case $\Delta$ expresses in terms of the slow roll parameters
as
\be\label{deltaSR}
\Delta = \frac{3}{2}-\nu= \eta_V-\epsilon_V +
\mathrm{higher~orders~in~slow~roll} = \frac12(n_S-1) - n_T
\ee
where
\be\label{indi}
n_S = 1 - 6 \, \epsilon_V + 2 \; \eta_V \quad , \quad  n_T = -2 \,
\epsilon_V = -r/8
\ee
are the scalar and tensor spectral indices, respectively and $r$
is the tensor to scalar ratio.
Thus, we can apply the results obtained in the previous sections
to this case in which the inflaton quantum fluctuations
decay into themselves via the trilinear coupling.
To leading order in slow roll this is done simply by
setting $\nu =\bnu = 3/2-(\eta_V-\epsilon_V)$ in the results
previously obtained. The slow roll parameters remain constant
to leading order in slow roll.

\vspace{2mm}

{\bf Superhorizon modes: $k=0$}

\vspace{2mm}

For superhorizon modes the results Eqs.(\ref{Gamma_2}) and
(\ref{resumX0ren}) show that the amplitude of the quantum
fluctuations decay as $\eta \rightarrow 0$ with the power law
given by Eq.(\ref{resumX0ren}) with $\beta_+ = 2
+\epsilon_V-\eta_V, \; \beta_- = -1 + \eta_V-\epsilon_V$ and
\be\label{gamai} \Gamma = \frac{8}{9} \left[\frac{g}{\pi \, H \,
(\eta_V-\epsilon_V)} \right]^2
\left[1+\mathcal{O}(\epsilon_V,\eta_V)\right] \;. \ee The
effective dimensionless coupling $g/H$  [see Eq.(\ref{coup})] is
related to the scale of inflation and the slow roll parameters:
\be \frac{g}{H}= \frac{V'''[\Phi_0]}{2 \; H}\sim
\frac{\xi_V}{\sqrt{2\; \epsilon_V}} \; \frac{3\; H}{2 \; M_{Pl}}
\ee To lowest order in slow roll  the power spectrum of curvature
perturbations $\triangle^2_{\mathcal{R}}$ is given by
\cite{liddle} \be\label{curva} \triangle^2_{\mathcal{R}}=
\frac{H^2}{8 \; \pi^2 \;  M^2_{Pl}  \; \epsilon_V   } =
\frac{1}{12 \; \pi^2 \;  M^6_{Pl}} \frac{V^3}{V'^2} \; . \ee This
allows to relate the effective dimensionless coupling $g/H$ to
quantities that are observable from CMB data: \be \label{goverH}
\frac{g}{H} = 3 \; \pi \;  \xi_V \left(
\triangle^2_{\mathcal{R}}\right)^{\frac{1}{2}}
\left[1+\mathcal{O}(\eta_V,\epsilon_V)\right]\; . \ee Then, we can
write $ \Gamma $ [Eq.(\ref{gamai})] completely in terms of slow
roll parameters and the power spectrum of curvature perturbations,
\be\label{GSR} \Gamma = \frac{8 \; \xi^2_V \;
\triangle^2_{\mathcal{R}}}{(\epsilon_V-\eta_V)^2}
\left[1+\mathcal{O}(\epsilon_V,\eta_V)\right] \; , \ee and in
terms of the scalar index and the tensor/scalar rate we have using
Eq. (\ref{indi}), \be\label{GSR2} \Gamma = 32 \; \frac{\xi^2_V
\;\triangle^2_{\mathcal{R}}}{(n_S - 1 +\frac{r}{4})^2}
\left[1+\mathcal{O}(\epsilon_V,\eta_V)\right] \; . \ee \noindent
Of particular importance is the behavior of the \emph{growing}
mode of the inflaton which gives the dominant term for $ \eta \to
0 $ after the wavelengths of the fluctuations cross the horizon.
>From Eqs. (\ref{X00}), (\ref{resumX0ren}) and (\ref{nusr}),  it is
given by \be \eta^{\beta_-+\Gamma} = \frac{\eta^{\eta_V-\epsilon_V
+\Gamma}}{\eta} \label{growmode} \;. \ee We see that the decay
constant $\Gamma$ acts as an anomalous scaling dimension for the
growing mode of the superhorizon fluctuations of the inflaton. The
decay rate $\Gamma$ slows down the growth of the dominant mode for
$ \eta \to 0 $ and fastens the decrease of the subdominant modes.

\vspace{2mm}

{\bf Modes inside the Hubble radius during slow roll inflation
$|k\eta| \gg 1$:}

\vspace{2mm}

The decay law of the quantum  fluctuations of the inflaton with
wavelengths deep inside the Hubble radius $|k\eta| =
k_{ph}(\eta)/H \gg 1$ is given by Eqs.(\ref{solu}) with \be
\label{USR} \Gamma(|k\eta|\gg 1) = \frac{g^2}{32 \; \pi  \; H^2}
\; \frac{H}{k_{ph}(\eta)} \; \left[1+\frac{2 \; H}{\pi \;
k_{ph}(\eta)}\left(\frac{1}{\eta_V-\epsilon_V}+
\ln\frac{k_{ph}(\eta)}{H}\right)\right] ~~ , ~~ k_{ph}(\eta)=
\frac{k}{C(\eta)}  ~~ , ~~ C(\eta) = -\frac{1 + \epsilon_V}{H \;
\eta} \; . \ee The term that is inversely proportional to the
slow-roll parameters and the logarithm of $ k_{ph}(\eta)/H $ are a
consequence of the almost collinear emission of superhorizon
quanta. This process corresponds to one of the decay lines in fig.
\ref{fig:inflatondecay} carrying small momentum with superhorizon
wavelengths as explained in the previous section. This process is
identified as the emission of
 \emph{bremsstrahlung radiation} of superhorizon quanta.

 Which of the terms in the bracket in  Eq.(\ref{USR}) dominates depends not
only on the numerical value of the slow roll parameters but also on how
 close the physical wave vector is to  horizon crossing. For
 very short wavelength modes, namely
 \be
k_{ph}(\eta) \gg \frac{2 \; H}{\eta_V-\epsilon_V}
\ee
 \noindent the second term is negligible and the result
 is similar to the inflaton decaying into massless
 conformally coupled particles. This is of course due to the
 fact that in this kinematic region the modes in the internal
 propagators are simply plane waves with Bunch-Davies initial
 conditions. On the contrary, as $ k_{ph}(\eta) $ approaches horizon crossing,
 \be
k_{ph}(\eta) \lesssim \frac{H}{\eta_V-\epsilon_V}\;,
\ee
 \noindent the emission of ultrasoft collinear quanta, namely
superhorizon \emph{bremsstrahlung radiation},
 becomes the dominant decay channel and the second term dominates.
 This crossover phenomenom can be interpreted as the phase space for
 collinear emission opening up near  horizon crossing.
This is because the phase space factor $1/k$ in
 eq. (\ref{ang}) is  effectively  $1/|k\eta|= H/k_{ph}(\eta)$
by dimensional reasons.

Thus, in slow roll, there is a wide region of physical momenta
\be
H \ll k_{ph}(\eta) \ll \frac{H}{\eta_V-\epsilon_V}  \; ,
\ee
for which  the first term in  the bracket in Eq.(\ref{USR}) can be neglected and the
leading slow-roll result  Eq.(\ref{goverH}) for $g^2/H^2$ can be used:
\be
\Gamma(|k\eta|\gg 1)=\frac{9}{16} \;
\left(\frac{\xi_V \; \triangle_{\mathcal{R}}}{k \, \eta}\right)^2
(1 + 2 \, \epsilon_V)\left\{\frac{1}{\eta_V- \epsilon_V}+
\log[k\,\eta(1-\epsilon_V)]\right\} \; .
\ee

\section{Implications for the decay of scalar and tensor perturbations and
non-gaussianity}

While we have focused on the decay of the  inflaton quantum fluctuations
during slow roll we can extrapolate our
results to see  how our findings \emph{may} provide
corrections to the power spectra of scalar and tensor perturbations.

\subsection{Curvature perturbations:}
For scalar perturbations, the action for the  gauge invariant perturbation
\be
u_k = C(\eta) \; \phi_k + \frac{\Phi'_0}{H} \; \psi_k \; ,
\ee
\noindent has a
simple form at quadratic order \cite{bran} and obeys the equation of motion
\be
u''_k + \left[k^2 - \frac{z''}{z}\right]u_k =0 ~~;~~ z \equiv
\frac{\Phi'_0}{H}  \; .
\ee
Here $\Phi_0$ is the inflaton expectation value, $ \phi_k $ the inflaton
fluctuation and  $\psi_k$ are the spatial curvature perturbations.
The gauge invariant perturbation $u_k$ is related to the curvature
perturbation on comoving hypersurfaces $\mathcal{R}$
as \cite{lidsey,liddle,bran} $ u_k= -z  \; \mathcal{R}_k $.
 Therefore, the power spectrum of the curvature perturbation
 $\triangle^2_{\mathcal{R}}$ is directly related to the corresponding
spectrum  of the gauge invariant perturbation $u_k$ \cite{lidsey,liddle,bran}
 \be\label{powspec}
\triangle^2_{\mathcal{R}} = \frac{k^3}{2\pi^2} \left|\frac{u_k}{z}
 \right|^2 \; .
\ee
For superhorizon modes,  $k^2 \ll z''/z $, the only relevant contribution is
the growing mode $ u_k= A_k \; z $.
Therefore, well after horizon crossing the power spectrum of the
curvature perturbation is constant in time and given by
\be \label{afterPS}
\triangle^2_{\mathcal{R}} = \frac{k^3}{2\pi^2}
\left|A_k  \right|^2 \; .
\ee
 During slow roll \cite{lidsey,liddle}
 \be
\frac{z''}{z}= 2 \;  C^2(\eta) \; H^2 \left[ 1 + \epsilon_V - \frac32 \; \eta_H
\right] =
 \frac{2}{\eta^2}\left[1+\frac{3}{2}(\eta_V+\epsilon_V)+
\mathcal{O}(\eta^2_V,\epsilon^2_V,\eta_V\epsilon_V,\xi_V)\right]
 \equiv \frac{\nu^2-\frac{1}{4}}{\eta^2}
\ee
\noindent where we used Eqs.(\ref{etav}) and (\ref{quasiDS}) and which implies
 \be
\nu =  \frac{3}{2}+\epsilon_V+\eta_V+
\mathcal{O}(\eta^2_V,\epsilon^2_V,\eta_V\epsilon_V,\xi_V)
\ee
Therefore, the small parameter $\Delta$ for the gauge invariant perturbation
is given by
\be\label{Dig}
\Delta=-\epsilon_V-\eta_V+
\mathcal{O}(\eta^2_V,\epsilon^2_V,\eta_V\epsilon_V,\xi_V)
\ee
Without interactions among fluctuations (no decay of fluctuations),
the growing mode for superhorizon wavelengths
to lowest order in slow roll is given by
\be \label{desiu}
u_k(\eta) = A_k ~\eta^{-(1+\epsilon_V+\eta_V)}
\ee
When a cubic interaction for the perturbation $ u_k(\eta) $ is
introduced, the results of the previous sections imply that
an anomalous scaling dimension, namely the decay rate $ \Gamma$
 will appear in Eq. (\ref{desiu}), i. e.
$$
u_k(\eta) = A_k ~\eta^{-(1+\epsilon_V+\eta_V-\Gamma )} \; ,
$$
as in Eq. (\ref{growmode}) for the inflaton fluctuations.
Unless $z$ acquires the \emph{same} anomalous dimensions, the
amplitude of the power spectrum [Eq. (\ref{powspec})] will depend on time.

\medskip

In order to study the decay of \emph{curvature perturbations} the
next step in this program is to obtain the cubic vertex for the
variable $u$ and to compute the one-loop self energy. This will
ensure the gauge invariance of the results. The gauge
invariant formulation of ref.\cite{bran} has to be extended
to higher order of perturbations, as for example in ref.\cite{bart}, or
alternatively, we can work in a fixed gauge. The cubic interaction
vertex for three scalars has been computed in ref.\cite{maldacena}.
In particular, the contribution to the self-energy from ghost loops must be
included.

The computation of the self-energy corrections will follow the same
lines presented in the previous sections with the extra feature
of momentum dependent vertices and ghost-loops. The infrared
behavior of the loops will be regulated by $\Delta$ given by Eq.(\ref{Dig}).

\medskip

In order to obtain corrections for the power spectrum of curvature
perturbations, the secular terms arising from
the equations of motion will have to be resummed by
DRG for the whole range of physical
momenta, until and beyond horizon crossing, and establish the behavior
of the growing mode. At least two sources of corrections to the
index of the power spectra may be expected:

i) from the amplitude of the growing mode $ A_k $, which is obtained by
matching the solutions deep within the Hubble radius and well after horizon
  crossing,

ii) from the anomalous scaling dimension of the growing mode,
namely the decay rate  $\Gamma$.

\medskip

Furthermore, in order to extract the power spectrum, the
  corrections to the dynamics of the zero mode of the inflaton
  $\Phi_0$ from the coupling of the zero mode to the fluctuations
  must be obtained. This study will indicate whether the variable
  $z$ also acquires an anomalous dimension and if so, whether the
  ratio $u_k/z$ for the growing mode is time independent when the decay
  process is accounted for. This study is in progress.

  \subsection{Gravitational waves:}
  For tensor modes, i.e, gravitational waves the action at
  the quadratic level is simple \cite{bran} and the field operators for the
  (gauge invariant) gravitational waves are expanded in terms of the mode
  functions $V_k$ which satisfy
  \be
V''_k + \left[k^2 - \frac{C''}{C}\right]V_k=0 \; ,
\ee
  \noindent where $\frac{C''}{C}$ in given by Eq. (\ref{c2c2}).
We then have
  \be
\frac{C''}{C} =\frac{\nu^2-\frac14}{\eta^2}+
\mathrm{higher~orders~in~slow~roll} \; ,
  \ee
i. e. ,
$$ \nu = \frac32 + \epsilon_V +\mathrm{higher~orders~in~slow~roll}\quad , \quad
 \Delta = -\epsilon_V +\mathrm{higher~orders~in~slow~roll} \; .
$$
  Thus, during quasi de Sitter slow roll, the mode functions for gravitational
waves are Hankel functions with an index $\nu$ slightly different from $3/2$.

\medskip

   The three graviton vertex was computed in
   refs.\cite{IRcosmo,maldacena}. The Born
   scattering amplitude for three gravitons in a de Sitter
space-time features both infrared as well as secular divergences\cite{IRcosmo}.
   These infrared divergences precisely arise because the mode
   functions for gravitons in de Sitter space-time are Hankel
   functions with index $\nu=3/2$.
The remaining long time ($\eta \rightarrow 0$) divergences can be
consider as secular terms akin to those found above.

    During quasi de Sitter slow roll inflation, the parameter $\Delta$
   can  regulate the infrared divergences found in
   refs. \cite{IRcosmo,dolgov}, and the DRG implemented here
will provide a resummation of the secular long time divergences found in
refs.\cite{IRcosmo} and \cite{dolgov}.

\medskip

   While the three graviton scattering amplitude has been computed
   in the Born approximation, the {\it full self-energy} for gravitons has
   {\it not yet} been obtained. In order to understand the decay law for
   gravitons, the program presented in this article must be
   implemented. Such program for gravitons, as in the case of the scalar
   perturbations discussed above, may require including the
   contribution from ghost loops to the graviton self-energy.
  Clearly, carrying this program in either case is a  task
beyond the scope of this article.

\subsection{Estimates of the cubic coupling and the decay rate
from the WMAP data}
 In the slow roll approximation the slow-roll parameters
themselves are slowly varying functions of time, in particular
\be\label{etadot}
\frac{\dot{\eta}_V}{H} = 2 \; \epsilon_V \; \eta_V-
\sqrt{2\epsilon_V} \; \frac{2M_{Pl}}{3H}  \; \frac{g}{H}+
\mathcal{O}(g \; \epsilon_V^{3/2}) \; .
\ee
The slow roll approximation entails that $\dot{\eta}_V/H \ll 1$
which in turn requires
\be\label{weakcond}
\frac{g}{H} \ll \frac{3
H}{2\sqrt{2\epsilon_V}\,M_{Pl}} \; .
\ee
WMAP gives the following value for $\triangle^2_{\mathcal{R}}$ \cite{peiris}
\be
\triangle^2_{\mathcal{R}} \simeq 2.2 \times 10^{-9}
\ee
\noindent which when combined with Eq.(\ref{curva}) yields the
following estimate on the scale of inflation
 \be\label{HWMAP}
H \sim 10^{15} \;  \sqrt{\epsilon_V} \;  [\mathrm{Gev}] \ee

The WMAP data on $d\,n_s/d\ln k$ \cite{peiris} suggests that
 $\xi_V \simeq 0.028 \pm 0.015$ which combined with Eq.
 (\ref{goverH})  leads to the following estimate on the cubic coupling,
 \be\label{gs}
\frac{g}{H} \simeq 1.3 \times 10^{-5}
\ee which places the strength of
 the dimensionless coupling within the validity of the slow roll
 approximation and perturbation theory. These results  in turn lead to the
following estimate for the rate $\Gamma$ [Eq. (\ref{GSR})],
\be
3 \times 10^{-8} \gtrsim \Gamma \gtrsim 3.6 \times 10^{-9} \; .
\ee
This gives in cosmic time for a
typical value $ H \simeq 10^{14}$GeV
\be
10^7 \; GeV \gtrsim \Gamma_{dS} = H \; \Gamma \gtrsim 10^6 \; GeV \; .
\ee

\subsection{Connection with non-gaussianity:}
Non-gaussianity of the spectrum of fluctuations is associated with
three (and higher) point correlation functions. An early assessment
of non-gaussian features  of temperature fluctuations in an
interacting  field theory  was given in ref.\cite{allen}. In
ref.\cite{srednicki} the simplest inflationary potential with a
cubic self-interaction for the inflaton field was proposed as a
prototype theory to study possible departures from gaussianity.
The three point correlation function of a scalar field in a theory with
cubic interaction as well as the four point correlation function in
a theory with quartic interaction were calculated in
ref.\cite{uzan,bartolo}.

\begin{figure}[ht!]
\begin{center}
\includegraphics[height=2.5in,width=2.5in,keepaspectratio=true]{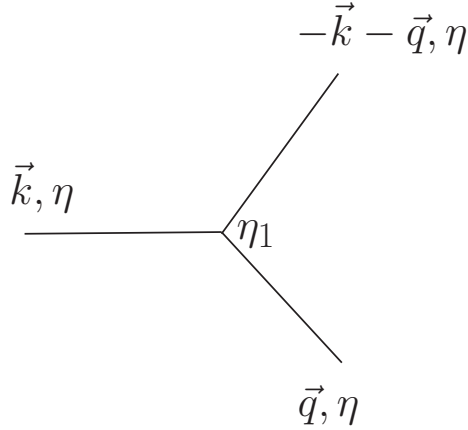}
\caption{Equal time three point function $\langle \chi(\vk,\eta)
\; \chi(\vq,\eta) \; \chi(-\vk-\vq,\eta)\rangle$ in the Born
approximation. The time coordinate $\eta_1$ of the vertex is
integrated.} \label{fig:expecval}
\end{center}
\end{figure}

The long time ($\eta\rightarrow 0$) behavior of the equal time
three point correlation function in the scalar field theory defined by
Eq.(\ref{1field})  for $M = 0$ (and hence $\nu = \frac32$),
is given by \cite{srednicki}
\be \label{3point}\langle
\chi(\vk,\eta) \; \chi(\vq,\eta) \; \chi(-\vk-\vq,\eta)\rangle =
\frac{2\pi^3}{3} \; C^3(\eta) \; g \;  H^2
\frac{F(\vk,\vq;\eta)}{\left[ k \;  q \;  |\vk+\vq| \right]^3}
\ee
\noindent where
\be \label{ef}
F(\vk,\vq;\eta)= \left[k^3+q^3+|\vk+\vq|^3
\right]\left[\ln (k_T\eta) +\gamma\right]-(k^2+q^2+|\vk+\vq|^2
)k_T+k\,q\,|\vk+\vq|~~;~~k_T= k+q+|\vk+\vq|
\ee
A diagrammatic interpretation of the equal time expectation
value Eq.(\ref{3point}) is depicted  in fig. \ref{fig:expecval}, which
illustrates the {\it similarity} with the {\it decay process} depicted in fig.
\ref{fig:inflatondecay}.

Furthermore, the logarithmic secular term in  Eq.(\ref{ef}) indicates
that the three point function features
secular divergences {\bf even at the tree level}. It is argued in
refs. \cite{srednicki,uzan,bartolo}
that $ -\ln [k_T\eta]\sim 60$  which is the
number of e-folds from the time when fluctuations of wavenumber
$k_T$ first crossed the horizon till the end of inflation. However,
such infrared logarithms are {\bf secular terms} and have precisely
the {\bf same} origin as in the self-energy kernel and in the inflaton
fluctuations discussed here (secs. IIIC and IIID). The same holds for the
 infrared logarithms in the three graviton scattering
vertex \cite{IRcosmo,dolgov}.

In particular, the self-energy computation
corresponds to a further integration over the loop momentum $q$. In a
fairly loose manner, the self-energy is basically the square of the
three point correlation function integrated over the loop momentum.
This is akin to the unitarity relation between the imaginary part of
the forward scattering amplitude and the square of the transition
amplitude in S-matrix theory.

In summary, the {\it interaction} between the fluctuations gives rise
to  {\it non-gaussian} correlations which are determined by the three
point function which is  precisely related to the
{\it self-energy}  and  the {\it decay} of the quantum fluctuations.
Therefore, the decay of the quantum fluctuations of the scalar field
will also lead to non-gaussian correlations and
non-gaussianity in the power spectrum.

The direct relationship between the self-energy, decay and
non-gaussian features of the power spectrum will be the subject of
further study.

\section{Conclusions and further questions}

In this article we have studied particle decay of fields
minimally coupled to gravity in the case when the mass of the fields
is $\ll H$ during inflation. Unlike the  decay into  massless
fields conformally coupled to gravity, this case features a strong
infrared behavior which leads to novel results.

We have implemented the dynamical renormalization group resummation
program introduced in ref.\cite{desiterI} combined with an expansion
in a small parameter $\Delta$ which regulates the infrared.

In the case of exact de Sitter inflation, $\Delta$ is a constant equal to
the ratio of the mass of the decay products to the Hubble constant,
while in slow roll inflation $\Delta$ is a simple function of slow
roll parameters. The expansion in $\Delta$ is akin to the
$\varepsilon$ expansion in critical phenomena in dimensional
regularization. The dynamical renormalization group provides a
resummation of the long-time secular divergences which determine the
decay law of quantum fluctuations.

The lack of energy conservation in an expanding
cosmology leads to the lack of kinematic
thresholds for particle decay. In particular, this possibility leads to the
\emph{self-decay} of quantum fluctuations whenever a
self-interaction is present.

We have studied the decay of a particle for a  cubic selfcoupled
scalar field in de Sitter space-time and applied the results to the
\emph{self-decay} of the inflaton quantum fluctuations
during quasi de Sitter, slow roll inflation.
We focused on extracting the decay law both for
wavelengths well inside and well outside the Hubble radius. In both
cases the strong infrared behavior enhances the decay.

The decay of fluctuations with wavelengths much smaller than the
Hubble radius {\bf is enhanced} by the collinear emission of
ultrasoft quanta, this process is identified as
 \emph{bremsstrahlung radiation} of superhorizon quanta. As the
 physical wavelength approaches the horizon, the phase space for
 this process opens up becoming the dominant decay channel for short
 wavelength modes in the region
 \be
H \ll k_{ph}(\eta) \lesssim \frac{H}{\eta_V-\epsilon_V} \; .
\ee
The decay of short wavelength modes hastens as the physical
wavelength approaches the horizon as a consequence of the opening up
of the phase space.

Superhorizon quantum fluctuations decay as a power law $\sim
\eta^{\Gamma}$ in conformal time, where $\Gamma$ is determined by
the following combination of the slow roll parameters and the
amplitude of curvature perturbations \be \Gamma = \frac{32 \;
\xi^2_V \; \triangle^2_{\mathcal{R}}}{(n_s-1+\frac{r}{4})^2}
\left[1+\mathcal{O}(\epsilon_V,\eta_V)\right]
 \ee
This decay law entails that the growing mode for superhorizon
wavelengths evolves as $  {\eta^{\eta_V-\epsilon_V +\Gamma}}/{\eta}
$ hence $\Gamma$ provides an {\it anomalous scaling dimension}
slowing down the growing mode for $ \eta \to 0 $.

The recent WMAP data indicate that $ 3. \times 10^{-8}  \gtrsim
\Gamma \gtrsim 3.6 \times 10^{-9} $. This corresponds to  a decay rate in
cosmic time $ 10^7 GeV \gtrsim \Gamma_{dS} \equiv H \; \Gamma \gtrsim 10^6 $GeV.
Although these values may seem small, it must be noticed that the
decay is a secular, namely cumulative effect.

We discussed some potential applications and implications for
primordial scalar and tensor perturbations as well as the
relationship between the decay processes studied in this article and
the generation of non-gaussian features in the primordial power
spectrum.

The results of our study bring about several questions:

\begin{itemize}
\item{The generation of superhorizon fluctuations during inflation is
usually referred to as `acausal'. However, we
have found that fluctuation modes deep inside the horizon decay into
superhorizon modes, therefore there is a coupling between modes
inside and outside the horizon. The phase space for this process
opens up as the physical wavelength approaches the horizon. It is
natural to conjecture that this process that couples modes inside
and outside the horizon with a coupling that effectively depends on
the wave vector, will ultimately lead to {\it distortions} in the power
spectrum. This distortion will necessarily be small in slow-roll
since the coupling is of the order of the slow roll parameters, but it
{\it may compete} with the running of the spectral index from the
non-interacting theory which is itself of quadratic order in slow
roll.  }

\item{In the non-interacting theory, the equation of motion for the gauge
invariant Newtonian potential (equal to curvature
perturbation) features a constant of motion for superhorizon
wavelengths \cite{bardeen,bran}. This is used to
estimate the spectrum of density perturbations in inflationary
universe models. It is conceivable
that this conservation law will no longer hold in higher order in
slow roll when interactions are included.
We expect this to be the case for two reasons: the
coupling between modes inside and outside the horizon as well as the
decay of superhorizon modes. Clearly the violation of the
conservation law, if present, will be small in slow-roll, but this
non-conservation may also lead to {\it distortions} in the power spectrum.
}

\item{While we have focused on the decay process during inflation,
our results, in particular the decay of superhorizon
fluctuations and the coupling between modes inside and outside the
Hubble radius, raise the possibility of similar processes being
available during the radiation dominated phase. If this is the case,
the decay of short wavelength modes into superhorizon modes can
serve as an active process for seeding superhorizon fluctuations.  }

\end{itemize}

Forthcoming  observations of CMB anisotropies as well as large
scale surveys with ever greater precision will provide a
substantial body of high precision observational data which may
hint at corrections to  the generic and robust predictions of
inflation. If such is the case these observations will  pave the
way for a better determination of inflationary scenarios. Studying
the possible observational consequences of the quantum phenomena
found in this article will therefore prove a worthwhile endeavor.

\begin{acknowledgments} D.B.\ thanks the US NSF for support under
grant PHY-0242134,  and the Observatoire de Paris and LERMA for
hospitality during this work. This work is supported in part by the
Conseil Scientifique de l'Observatoire de Paris through an `Action
Initiative'.
\end{acknowledgments}

\appendix
\section{Self-energy kernel for $\vk =0$}

We compute here the kernel $\mathcal{K}_{\bnu}(0,\eta,\eta')$
\be\label{K0A}
\mathcal{K}_{\bnu}(0;\eta,\eta') = \frac{1}{\pi^2}
\int_0^{\infty} q^2  \; dq ~
\mathrm{Im}\left\{\left[S_{\bnu}(q,\eta)S^*_{\bnu}(q,\eta')
\right]^2\right\} \quad ,  \quad \bnu=\frac{3}{2}-\Delta \; .
\ee
The infrared divergences at $ \Delta = \frac{3}{2} - \bnu = 0 $
arise from the small momenta behavior of the integrand in Eq.(\ref{K0A}).
To extract such behavior it is convenient to write the Hankel function
$H^{(2)}_{\bnu}(q\eta)$ in Eq. (\ref{Snu}) as \cite{grad}
\be\label{Hankel}
H^{(2)}_{\bnu}(z) = \frac{i}{\sin \pi \bnu} \left[  J_{-\bnu}(z)
- e^{i \; \pi \; \bnu} \; J_{\bnu}(z) \right] \; .
\ee
\noindent The leading $ q \to 0 $  behavior follows from the Bessel functions
$J_{-\bnu}(z)$ since \cite{grad},
$$
J_{-\bnu}(z) \buildrel{z \to 0}\over=  \left(\frac{2}{z}\right)^{\bnu} \;
\frac{1}{\Gamma(1-\bnu)}\left[1 + \mathcal{O}(z^2)\right] \; .
$$
We find upon taking the imaginary part,
\be\label{IRdomi}
\frac{q^2}{\pi^2}\mathrm{Im}\left\{\left[S_{\bnu}(q,\eta)
 \; S^*_{\bnu}(q,\eta')\right]^2\right\}\buildrel{q \to 0}\over=
 -\left(\frac{4}{q^2 \; \eta \, \eta'}\right)^{\bnu-1} \;
\frac{ \Gamma^2(\bnu)}{2 \; \pi^3 \; \bnu }
\left[\left(\frac{\eta}{\eta'}
\right)^{\bnu}-\left(\frac{\eta'}{\eta} \right)^{\bnu}
\right]\left[1+\mathcal{O}(q^2)\right] \; .
\ee
The behaviour $ q^{2 - 2 \,\bnu} = q^{2 \, \Delta - 1} $  for $ q \to 0 $
in the integrand of  Eq. (\ref{K0A})
implies a simple pole at $ \Delta = 0 $ with a $ \eta$-independent residue
\cite{gel}. It is then convenient to add and to subtract from Eq. (\ref{K0A})
the low $q$ behaviour Eq. (\ref{IRdomi}). We find,
\be\label{K0split}
\mathcal{K}_{\bnu}(0;\eta,\eta')=
I_1(\eta,\eta';\mu)+I_2(\eta,\eta';\mu)
\ee
\noindent with the integrals $I_{1,2}$ given by
\bea
&&I_1(\eta,\eta';\mu)  =  \int_0^{\mu} dq \left(
\frac{q^2}{\pi^2}
\mathrm{Im}
\left\{\left[S_{\bnu}(q,\eta) \; S^*_{\bnu}(q,\eta')\right]^2\right\}+
\left(\frac{2}{q^2 \; \eta \, \eta'}\right)^{\bnu-1} \;
\frac{ \Gamma^3(\bnu)}{2 \; \pi^3 \; \Gamma(1+\bnu)}
\left[\left(\frac{\eta}{\eta'}
\right)^{\bnu} -\left(\frac{\eta'}{\eta} \right)^{\bnu}
\right] \right) \cr \cr
&&- \left(\frac{2}{\eta \, \eta'}\right)^{\bnu-1} \;
\frac{ \Gamma^3(\bnu)}{2 \; \pi^3 \; \Gamma(1+\bnu)}
\left[\left(\frac{\eta}{\eta'}
\right)^{\bnu}-\left(\frac{\eta'}{\eta} \right)^{\bnu}
\right] \int_0^{\mu} dq \; q^{2 - 2 \, \bnu} \; ,
\label{I1} \\
&&I_2(\eta,\eta';\mu) =  \frac{1}{\pi^2} \int_{\mu}^{\infty} q^2 \;  dq~
\mathrm{Im}\left\{\left[S_{\bnu}(q,\eta)S^*_{\bnu}(q,\eta')\right]^2\right\}
\label{I2} \; .
\eea
Here, $ \mu > 0 $ is an arbitrary parameter temporarily introduced to split
the integrals. The sum $ I_1(\eta,\eta';\mu)+I_2(\eta,\eta';\mu) $ is clearly
$\mu$-independent.
The second integral in $I_1(\eta,\eta';\mu)$  [Eq. (\ref{I1})]
can be done straightforwardly,
\be
\int_0^{\mu} q^{2-2\bnu} \; dq = \frac{\mu^{2 \; \Delta}}{2 \; \Delta}\, .
\ee
\noindent This simple integral clearly displays the infrared
divergence and the origin of the pole in $\Delta$: the emission of
superhorizon quanta for which $q\eta,q\eta' \ll1 $ as depicted in
fig. \ref{fig:fivfi}.

Keeping the pole in $\Delta$ plus the
leading logarithmic terms, and neglecting higher order terms in
$\Delta$, we find for $\mu \eta,\mu \eta'  \ll 1 $
\be\label{I1lead}
I_1(\eta,\eta';\mu) = \frac{1}{6\pi^2}
\frac{1}{2\Delta} \left[\frac{\eta'}{\eta^2}-
\frac{\eta}{\eta^{'2}} \right]+ \frac{1}{6\pi^2} \left[
\frac{\eta'}{\eta^2} \ln[-\mu \eta]-\frac{\eta}{\eta^{'2}}\ln[-\mu
\eta']\right]+\mathcal{O}(\Delta)
\ee
We can set $3/2 -\bnu=\Delta=0 $ in
the integral $I_2$ since it is  infrared finite.
After lengthy but straightforward algebra, we find for
$\mu \eta,\mu \eta' \ll 1 $
\be\label{I2lead}
I_2(\eta,\eta';\mu) = -\frac{1}{8\pi^2}
\mathcal{P}\left( \frac{1}{\eta-\eta'}\right)+ \frac{1}{6\pi^2}
\Bigg\{ \left(\frac{\eta}{\eta^{'2}}-\frac{\eta'}{\eta^2}
\right)\left[\ln[\mu(\eta-\eta')]+ \gamma+\frac{5}{3} \right]+
\left(\frac{1}{\eta'}-\frac{1}{\eta} \right)\Bigg\}
\ee
\noindent the principal value P is defined by Eq. (\ref{K12}).
The first term in Eq.(\ref{I2lead}) is the kernel for a massless conformally
coupled field ($\bnu = 1/2$)\cite{desiterI}.
It is clear from Eqs. (\ref{I1lead}) and (\ref{I2lead}) that the
$\mu$ dependence cancels out as it should be:
\be
\mathcal{K}_{\bnu}(0;\eta,\eta')=\mathcal{K}_{\frac{1}{2}}(0;\eta,\eta')+
\frac{1}{6\pi^2} \Bigg\{   \left[\frac{1}{2\Delta}+\frac23 \right]
\left(\frac{\eta'}{\eta^{2}}-\frac{\eta}{\eta^{'2}}\right)
- \frac{\eta'}{\eta^2}\ln\left(\frac{\eta'}{\eta}
 \right)+
\left(\frac{\eta}{\eta^{'2}}- \frac{\eta'}{\eta^2} \right)
\ln\left[1-\frac{\eta}{\eta'} \right]
+ \frac{1}{\eta'}-\frac{1}{\eta} \Bigg\}
\ee
\noindent where $ \mathcal{K}_{\frac{1}{2}}(0;\eta,\eta') $ is
defined by Eq. (\ref{K12}).

\section{Self-energy kernel for $\vk \neq 0$}

The self-energy kernel $\mathcal{K}_{\bnu}(k;\eta,\eta')$ for the general case
is given by Eq.(\ref{kernel}). As highlighted in the case of $\vk=0$, there are
infrared divergences for $\bnu = 3/2$ which we regulated with
the parameter $\Delta=3/2-\bnu $. We now compute
$\mathcal{K}_{\bnu}(k;\eta,\eta')$ keeping poles in $\Delta$ and the leading
logarithms in $\eta$. The strategy is to separate the regions in the loop
integral that contain the infrared divergences. Thanks to the azimuthal
invariance we can write
\be
\int d^3q = 2\pi \int_0^\infty q^2~dq \int^{+1}_{-1}d(\cos\theta)
\ee
\noindent with $\theta$ the angle between $\vec{q}$ and $\vk$.
Furthermore, we change from the integration variable $\theta$ to
$ p=|\vec{q}-\vk|$,
\be\label{ang}
 p= \sqrt{ q^2 +k^2 - 2 \, k \, q \, \cos\theta} \quad , \quad
d(\cos \theta ) = \frac{dp}{k \; q} \;,
\ee
\noindent which clearly displays the phase space factor $1/k$.
Eq.(\ref{kernel}) takes then the symmetric form
\be
\mathcal{K}_{\bnu}(k;\eta,\eta') = \frac{\eta \, \eta'}{32 \, k}
\int_0^{\infty} q \; dq \int_{|q-k|}^{q+k} p \; dp \;
\mathrm{Im} \left[  H^{(2)}_{\bnu}(q\eta)  \;  H^{(1)}_{\bnu}(q\eta') \;
H^{(2)}_{\bnu}(p\eta)\;  H^{(1)}_{\bnu}(p\eta') \right] \; .
\ee
\noindent where we used Eq. (\ref{Snu}).
The integral over $p$ can be performed using \cite{grad}
\be\label{iden}
\int p \; H^{(2)}_{\bnu}(p\eta) \; H^{(1)}_{\bnu}(p\eta') \, dp =
F_{\bnu}(p,\eta,\eta') \; ,
\ee
\noindent where,
\be\label{Fnu}
F_{\bnu}(p,\eta,\eta')= \frac{p}{\eta^2-\eta'^2} \;
\left[\eta' \;  H^{(2)}_{\bnu}(p\eta) \; H^{(1)}_{\bnu-1}(p\eta')-\eta
 \; H^{(2)}_{\bnu-1}(p\eta) \; H^{(1)}_{\bnu}(p\eta') \right] \; .
\ee
We obtain thus for the kernel,
\be\label{Kk}
\mathcal{K}_{\bnu}(k;\eta,\eta') = \frac{\eta \, \eta'}{32 \, k}
 \; \mathrm{Im} \int_0^{\infty} q \; dq
\left[ F_{\bnu}(q+k,\eta,\eta') -  F_{\bnu}(|q-k|,\eta,\eta')  \right]
 H^{(2)}_{\bnu}(q\eta)  \;  H^{(1)}_{\bnu}(q\eta') \; .
\ee
The infrared divergences at $ \Delta = \frac{3}{2} - \bnu = 0 $
arise from the $ q \to 0 $ and $ q \to k $
behavior of the integrand in Eq.(\ref{Kk}).
To extract such divergences we use the small argument behavior of
the Hankel functions \cite{grad}
\bea\label{HH}
&& H^{(2)}_{\bnu}(q\eta)  \;  H^{(1)}_{\bnu}(q\eta')  \buildrel{q \to 0}\over=
\frac{\Gamma^2(\bnu)}{\pi^2} \;
\left(\frac{4}{q^2 \; \eta \, \eta'} \right)^{\bnu}\left[ 1 + \mathcal{O}(q^2)
 \right] \; , \cr \cr
&&F_{\bnu}(q,\eta,\eta') \buildrel{q \to 0}\over= - \frac{\Gamma(\bnu) \;
\Gamma(\bnu-1)}{2 \; \pi^2} \;
\left(\frac{4}{\eta \, \eta'} \right)^{\bnu} \; q^{2 - 2 \,\bnu}
\left[ 1 + \mathcal{O}(q^2) \right]\; .
\eea
We see from Eqs.(\ref{Kk}) and (\ref{HH}) that the integrand in Eq.(\ref{Kk})
behaves as $ q^{2 - 2 \,\bnu} = q^{2 \, \Delta - 1} $ for $ q \to 0 $
and as $ |q-k|^{2 - 2 \,\bnu} = |q-k|^{2 \, \Delta - 1} $ for $ q \to k $.
Therefore, the kernel has a simple pole at $ \Delta = 0 $ \cite{gel}.
It is then convenient to add and to
subtract from Eq. (\ref{Kk}) the behaviour for $ q \to 0 $ and $ q \to k $
Eq. (\ref{HH}) as we did in the Appendix A for the $k=0$ case. We find,
\bea\label{Kk2}
&&\mathcal{K}_{\bnu}(k;\eta,\eta') = \frac{\eta \, \eta'}{32 \, k}
\; \mathrm{Im} \int_0^{\infty} dq \left\{ q \;
\left[ F_{\bnu}(q+k,\eta,\eta') -  F_{\bnu}(|q-k|,\eta,\eta')  \right]
 H^{(2)}_{\bnu}(q\eta)  \;  H^{(1)}_{\bnu}(q\eta')  \right. \cr \cr
&& - \left. \frac{\Gamma(\bnu) \; \Gamma(\bnu-1)}{2 \, \pi^2} \;
\left(\frac{4}{\eta \, \eta'} \right)^{\bnu} \; k \; H^{(2)}_{\bnu}(k\eta)
  \;  H^{(1)}_{\bnu}(k\eta') \left[4 \, (\bnu -1) \;  \theta(\mu - q) \;
q^{2-2\bnu} + \theta(k - q + \mu ) \; \theta(q - k + \mu ) \;
| q - k |^{2-2\bnu} \right] \right\} \cr \cr
&& +\frac{\Gamma(\bnu) \; \Gamma(\bnu-1)}{8 \, \pi^2} \; (\bnu - \frac12)  \;
\frac{\mu^{3-2\bnu}}{\frac32 -\bnu} \;
\left(\frac{4}{\eta \, \eta'} \right)^{\bnu-1} \;  k \; \mathrm{Im} \;
 H^{(2)}_{\bnu}(k\eta)   \;  H^{(1)}_{\bnu}(k\eta') \; .
\eea
$ \mu > 0 $ is an arbitrary parameter temporarily introduced as in Appendix A
and we used that
$$
 F_{\bnu}(q+k,\eta,\eta') -  F_{\bnu}(|q-k|,\eta,\eta')
\buildrel{q \to 0 }\over= 2 \, q
\frac{\partial  F_{\bnu}(k,\eta,\eta')}{\partial k} =  2 \, q \, k \;
 H^{(2)}_{\bnu}(k\eta)   \;  H^{(1)}_{\bnu}(k\eta') \; .
$$
$ \mathcal{K}_{\bnu}(k;\eta,\eta')$ is clearly $\mu$-independent
as one can easily check by computing the derivative with respect to $ \mu $
of the r. h. s. of eq.(\ref{Kk2}). Notice that $ \theta(\mu - q) $ is nonzero
for $ q < \mu $ while $ \theta(k - q + \mu ) \; \theta(q - k + \mu ) $
does not vanishes for $ k-\mu < q <  k+\mu$.
The pole at $ \frac32 -\bnu =\Delta = 0 $
is explicit in the last term of Eq.(\ref{Kk2}) while the integral over $q$
is finite for $ \bnu=\frac32$ and $ k \neq 0 $.

This analysis  for the self-energy kernel shows that all
infrared singularites emerge from the regions $|q\eta|,|q\eta'| \ll 1$
and  $|p\eta|,|p\eta'| \ll 1$ corresponding to the internal line in the loop
which carries momentum $q$ or $p=|q-k|$ being superhorizon.  Both
regions give a similar contribution because they are equivalent upon
re-routing of the loop momentum and Bose symmetry.
Thus the conclusion of this analysis is that the
leading contributions to the self-energy arise from the
\emph{collinear} emission of superhorizon quanta [since $\cos\theta = 1$, see
Eq. (\ref{ang})].

The calculation of the integral for $\mathcal{A}_{\bnu}(k;\eta,\eta')$
in Eq. (\ref{Kk2}) is straightforward albeit
lengthy. These are facilitated by the expression of the
Hankel functions $H^{(1,2)}_{\frac{3}{2}}(q\eta)$ in terms of
elementary functions. Dropping contributions of the order $ \Delta $
the kernel in Eq. (\ref{Kk2}) becomes for $\mu \ll k, \; |\mu \eta|,
|\mu \eta'|\ll 1 $,
\be\label{sep}
\mathcal{K}_{\bnu}(k;\eta,\eta')=\mathcal{K}_{\bnu}^<(k;\eta,\eta')+
\mathcal{K}_{\bnu}^>(k;\eta,\eta')\ee
\noindent with
\bea
&&\mathcal{K}_{\bnu}^<(k;\eta,\eta') =
\frac{\Gamma(\bnu) \; \Gamma(\bnu-1)}{8 \, \pi^2} \; (\bnu - \frac12)  \;
\frac{\mu^{3-2\bnu}}{\frac32 -\bnu} \;
\left(\frac{4}{\eta \, \eta'} \right)^{\bnu-1} \; \mathrm{Im} \;
 H^{(2)}_{\bnu}(k\eta)   \;  H^{(1)}_{\bnu}(k\eta')
\buildrel{\Delta \to 0 }\over= \cr \cr
&&=\frac{1}{4 \, \pi^2 \, k^3 \, (\eta \, \eta')^2}
\left\{ k(\eta-\eta') \cos[k(\eta-\eta')] - (1 + k^2 \, \eta \, \eta')
\sin[k(\eta-\eta')] \right\} \times\cr \cr
&& \left[ \frac{1}{\Delta}
+ 2 \gamma -3 + \log\left(4 \, \mu^2 \, \eta \, \eta' \right)
+  D(k \eta, k \eta') \right]
\eea
where,
$$
D(k \eta, k \eta') \equiv \frac{\pi}{2} \, k^3 \,  (\eta \, \eta')^{3/2} \;
\mathrm{Im} \left[  H^{(2)}_{3/2}(k\eta) \; \left.
\frac{\partial  H^{(1)}_{\nu}(k\eta)}{\partial \nu} \right|_{\nu = 3/2}
+ H^{(1)}_{3/2}(k\eta') \;
\left. \frac{\partial  H^{(2)}_{\nu}(k\eta)}{\partial \nu} \right|_{\nu = 3/2}
\right]
$$
and
\bea
&&\mathcal{K}_{\bnu}^>(k;\eta,\eta') \buildrel{\Delta \to 0 }\over=
\frac{\eta \, \eta'}{32 \, k}
\; \mathrm{Im} \int_0^{\infty} dq \left\{ q \;
\left[ F_{3/2}(q+k,\eta,\eta') -  F_{3/2}(|q-k|,\eta,\eta')  \right]
 H^{(2)}_{3/2}(q\eta)  \;  H^{(1)}_{3/2}(q\eta')  \right. \cr \cr
&& - \left. \frac{k}{2 \, \pi \; \left(\eta \, \eta' \right)^{3/2}} \;
H^{(2)}_{3/2}(k\eta)   \;  H^{(1)}_{3/2}(k\eta') \left[\frac{2}{q} \;
\theta(\mu - q)  + \theta(k - q + \mu ) \; \theta(q - k + \mu ) \;
\frac{1}{| q - k |} \right] \right\} \; ,
\eea
where,
$$
 F_{3/2}(q,\eta,\eta') = \frac{2}{\pi \sqrt{\eta \, \eta'}}
\; e^{iq(\eta' - \eta) } \left[\frac{i}{\eta - \eta'} -
\frac{1}{q \, \eta \, \eta'} \right] \; .
$$
and
$$
H^{(2)}_{3/2}(k\eta)   \;  H^{(1)}_{3/2}(k\eta') =
\frac{2}{\pi \; k^3 \; (\eta \, \eta')^{3/2}} \;  e^{ik(\eta' - \eta) }
\left[ 1 + k^2 \; \eta \, \eta' + i \, k ( \eta - \eta') \right] \; .
$$
The integral for the kernel takes then form,
\bea\label{kmay}
&&\mathcal{K}_{3/2}^>(k;\eta,\eta')=
\frac{\mathrm{Im}}{8 \, \pi^2 \, k \; \eta \, \eta'} \int_0^{\infty}
\frac{dq}{q^2} \left(
\left[q^2 \; \eta \, \eta' + i \, q \, (\eta - \eta')+ 1 \right]
 \times \right. \cr \cr
&&\left. \left\{  \frac{i}{\eta - \eta'} \left[ e^{i(2\,q+k) (\eta' - \eta)}
-  e^{i(q+|q-k|) (\eta' - \eta)}\right]
- \frac{1}{\eta \, \eta'} \left[\frac{e^{i(2\,q+k) (\eta' - \eta)}}{q+k}
- \frac{ e^{i(q+|q-k|) (\eta' - \eta)}}{|q-k|} \right] \right\} \right.
\cr \cr
&&\left. - \frac{e^{i k (\eta' - \eta)}}{k^2 \; \eta \, \eta'}\left[
k^2 \; \eta \, \eta' + i \, k \, (\eta - \eta')+ 1 \right]
\left[ \frac{2}{q} \; \theta(\mu - q) + \frac{1}{| q - k |} \;
\theta(k - q + \mu ) \; \theta(q - k + \mu ) \right] \right)
\eea
The remaining calculations are straightforward but tedious. Integrating
over $q$ in Eq.(\ref{kmay}) yields for $\mu \ll k, \; |\mu \eta|,
|\mu \eta'|\ll 1 $,
\bea\label{finK}
&&\mathcal{K}_{3/2}^>(k;\eta,\eta')=
-\frac{1}{8\pi^2} \mathcal{P}\frac{\cos\alpha}{\eta-\eta'}+\nonumber \\
&& +\frac{1}{4 \, \pi^2 \, k^3 (\eta \, \eta')^2}\left\{
\left[ (1 + k^2 \; \eta \, \eta')\sin \alpha - \alpha \, \cos\alpha \right]
\left[\mathrm{Ci}(2 \, \alpha) - 1 + \gamma + \log\left(2 \; \frac{\mu^2}{k^2}
\; \alpha \right) \right] \right. \cr \cr
&&\left. + 2 \; \sin \alpha - \mathrm{Si}(2 \, \alpha)
\left[ (1 + k^2 \; \eta \, \eta')\cos\alpha + \alpha \, \sin \alpha  \right]
 \right\}
\eea
\noindent where $ \alpha \equiv k(\eta - \eta')$ ,
Ci(z) and Si(z) are the cosine and sine integral
functions respectively \cite{grad}.
The first term in Eq. (\ref{finK}) is the kernel for conformally coupled
massless particles and the principal part prescription is given
by Eq. (\ref{K12}).

Keeping consistently the leading order terms
in $\Delta$, namely the pole plus finite parts, the dependence on
$\mu$  cancels out as it should and we find for the kernel to leading order in
$\Delta$
\bea\label{finKlead}
&&\mathcal{K}_{\bnu}(k;\eta,\eta')\buildrel{\bnu \to 3/2}\over=
-\frac{1}{8\pi^2} \mathcal{P}\frac{\cos\alpha}{\eta-\eta'}+\nonumber \\
&& +\frac{1}{4 \, \pi^2 \, k^3 (\eta \, \eta')^2}\left\{
\left[ \alpha \, \cos\alpha -(1 + k^2 \; \eta \, \eta')\sin \alpha \right]
\left[ \frac{1}{\Delta} + \gamma - 2 +
\log\frac{2 \, k \, \eta \, \eta'}{\eta - \eta'}
- \mathrm{Ci}(2 \, \alpha) +  D(k \eta, k \eta')  \right] \right.\cr \cr
&&\left. +  2 \; \sin \alpha -
\left[ (1 + k^2 \; \eta \, \eta')\cos\alpha + \alpha \, \sin \alpha  \right]
\mathrm{Si}(2 \, \alpha)  \right\} + \mathcal{O}(\Delta) \; .
\eea

\vspace{2mm}

{\bf Long wavelength  limit: $|k\eta|, |k\eta'| \ll1$}

\vspace{2mm}

The behavior of the kernel Eq. (\ref{finK}) in the limit when the
wave vector $k$ corresponds to superhorizon wavelengths easily follows from
Eq.(\ref{finKlead}). We use
$$
Ci[\alpha] \buildrel{\alpha \to 0}\over=
  \ln \alpha +\gamma+\mathcal{O}(\alpha^2)
\quad , \quad
Si[\alpha] \buildrel{\alpha \to 0}\over=  \alpha - \frac{\alpha^3}{18}
+ \mathcal{O}(\alpha^5)
$$Gathering all of these results and keeping the lowest order
contributions in the limit $|k\eta|,|k\eta'| \ll 1$ we find
Eq. (\ref{Kbarnu}) for the kernel in the long-wavelength limit.

\vspace{2mm}

{\bf Short wavelength  limit: $|k\eta|, |k\eta'| \gg 1$}

\vspace{2mm}

In this limit the wave function $S_{\bnu}(k,\eta)$ is just like the
Minkowski space time free field mode function for massless fields,
namely
\be S_{\bnu}(k,\eta) =
 \frac{e^{-ik\eta}}{\sqrt{2k}}. \ee
It is straightforward to obtain the  limit $|k\eta|,|k\eta'| \gg 1$
of the kernel  from Eq. (\ref{finK})
In the short wavelength limit the kernel simplifies to
\be \label{kernelhik}
\mathcal{K}_{\bnu}(k,\eta,\eta')
\buildrel{\bnu \to 3/2, \; k^2\eta\eta'\gg 1 }\over=
-\frac{1}{8\pi^2} \mathcal{P}\frac{\cos
k(\eta-\eta')}{\eta-\eta'}-\frac{1}{4 \, \pi^2\,  k\, \eta\, \eta'}
\Bigg\{\sin k(\eta-\eta')\left[\frac{1}{\Delta}+\tilde{\mathcal{C}}
+\log\frac{k \, \eta \,\eta'}{\eta-\eta'} \right]+
\frac{\pi}{2} \; \cos k(\eta-\eta') \Bigg\} \; ,
\ee
\noindent with the constant $\tilde{\mathcal{C}}$ given by
$ \tilde{\mathcal{C}}= \ln 2+\gamma-2 $.

\end{document}